\documentclass[%
 reprint,
superscriptaddress,
 amsmath,amssymb,
 aps,
floatfix,
]{revtex4-2}
\usepackage{cancel}
\usepackage[caption=false]{subfig}
\usepackage{graphicx}
\usepackage{dcolumn}
\usepackage{bm}
\usepackage[hidelinks]{hyperref}
\usepackage{xcolor}
\usepackage[normalem]{ulem}
\definecolor{ra}{rgb}{0.8, 0.0, 0.0}
 
\newtheorem{thm}{Theorem}
\begin{document}

\preprint{APS/123-QED}
\title{Quantum field-theoretic machine learning}
\author{Dimitrios Bachtis}
\email{dimitrios.bachtis@swansea.ac.uk}
\affiliation{Department of Mathematics,  Swansea University, Bay Campus, SA1 8EN, Swansea, Wales, UK}
\author{Gert Aarts}
\email{g.aarts@swansea.ac.uk}
\affiliation{Department of Physics, Swansea University, Singleton Campus, SA2 8PP, Swansea, Wales, UK}
\affiliation{European Centre for Theoretical Studies in Nuclear Physics and Related Areas (ECT*) \& Fondazione Bruno Kessler
Strada delle Tabarelle 286, 38123 Villazzano (TN), Italy }
\author{Biagio Lucini}
\email{b.lucini@swansea.ac.uk}
\affiliation{Department of Mathematics,  Swansea University, Bay Campus, SA1 8EN, Swansea, Wales, UK}%
\affiliation{Swansea Academy of Advanced Computing, Swansea University, Bay Campus, SA1 8EN, Swansea, Wales, UK}

\include{ms.bib}

\date{February 18, 2021}

\begin{abstract}

We derive machine learning algorithms from discretized Euclidean field theories, making inference and learning possible within dynamics described by quantum field theory. Specifically, we demonstrate that the $\phi^{4}$ scalar field theory satisfies the Hammersley-Clifford theorem, therefore recasting it as a machine learning algorithm within the mathematically rigorous framework of Markov random fields. We illustrate the concepts by minimizing an asymmetric distance between the probability distribution of the $\phi^{4}$ theory and that of target distributions, by quantifying the overlap of statistical ensembles between probability distributions and through reweighting to complex-valued actions with longer-range interactions. Neural network architectures are additionally derived from the $\phi^{4}$ theory which can be viewed as generalizations of conventional neural networks and applications are presented. We conclude by discussing how the proposal opens up a new research avenue, that of developing a mathematical and computational framework of machine learning within quantum field theory.
\end{abstract}

\maketitle

\section{\label{sec:level1}Introduction}

 Relativistic quantum fields~\citep{jean} are formulated on Minkowski space where intricate mathematical problems related to the hyperbolic geometry emerge. By recasting Minkowski space as Euclidean significant simplifications can be obtained for certain cases: The hyperbolic problems are transformed to be elliptic, the Poincar\'e group becomes the Euclidean group where a positive-definite scalar product emerges, noncommuting operators are expressed as random variables and causality is formulated as a Markov property.

Of high importance is the reverse direction: that of arriving at a quantum field in Minkowski space by constructing it from one in Euclidean space. To make such prospects attainable a rigorous mathematical framework for quantum fields had to be established, and a series of relevant contributions led to advances known as constructive quantum field theory~\citep{glimmcft,velo1973constructive,seiler}. A connection between probability theory and quantum field theory was then established when quantum fields were constructed from Euclidean fields that satisfy Markov properties~\citep{Nelson1973,NELSON197397}.

Recently, applications of deep learning \citep{GoodBengCour16}, a class of machine learning algorithms which are able to hierarchically extract abstract features in data, have emerged in the physical sciences \citep{Carleo_2019}, including in field theories \citep{PhysRevLett.125.121601,PhysRevD.97.094506,PhysRevD.100.011501,PhysRevE.102.053306,PhysRevD.102.054501,PhysRevD.101.094507,favoni2020lattice,nicoli2021estimation,PhysRevResearch.2.023369} and in the study of phase transitions \citep{vanNieuwenburg2017,Carrasquilla2017,GIANNETTI2019114639,bachtis2020adding,bachtis2020extending,PhysRevB.94.195105,doi:10.7566/JPSJ.86.063001}.  Insights on machine learning algorithms have been obtained from the perspective of statistical physics \citep{Agliari_2020, doi:10.1080/00018732.2016.1211393,Goldt_2020,Alberici2020,Agliari2018,Barra2017,Barra2018,Mezard2017,Barra2012}, particularly within the theory of spin glasses \citep{Mezard1987}, or in relation to Gaussian processes \citep{Halverson:2020trp,46760,g.2018gaussian,novak2019bayesian,garriga-alonso2018deep}.

 A notable case of these algorithms is the framework of Markov random fields~\citep{Koller}, which introduces Markov properties on a graph-based representation to encode probability distributions over high-dimensional spaces. As quantum field theory and probability theory are evidently connected analytically~\citep{NELSON197397}, and computational investigations of quantum fields are feasible through the framework of lattice field theory~\citep{PhysRevD.10.2445}, a new challenge is anticipated to emerge: namely that of investigating machine learning from the perspective of quantum fields.

In this manuscript, we derive machine learning algorithms from discretized Euclidean field theories, making inference and learning possible within dynamics described by quantum field theory. From the mathematical point of view, we explore if the $\phi^{4}$ scalar field theory on a square lattice satisfies the Hammersley-Clifford theorem, therefore recasting it as a Markov random field which can complete machine learning tasks. From the equivalent perspective of physics, we treat the $\phi^{4}$ scalar field theory as a system with inhomogeneous coupling constants and we search based on its dynamics, which comprise local interactions, for the optimal values of the coupling constants that are able to complete a machine learning task. Specifically we consider the minimization of an asymmetric distance between the probability distribution of the $\phi^{4}$ theory and that of target distributions. We also quantify the overlap of statistical ensembles between probability distributions and investigate if reweighting to the parameter space of complex-valued actions with longer-range interactions is possible by utilizing instead the probability distribution of the approximating local inhomogeneous action.

We then proceed to derive neural network architectures from the $\phi^{4}$ scalar field theory which can progressively extract features of increased abstraction in data. We explore the implications of including a local symmetry-breaking term in the $\phi^{4}$ Markov random field, and rearrange the lattice topology to derive a $\phi^{4}$ neural network which can be viewed as a generalization of conventional neural network architectures. Based on the equivalence between the $\phi^{4}$ scalar field theory and the Ising model under a certain limit, we discuss how the $\phi^{4}$ neural network can provide novel physical insights to the interpretability of a notable class of machine learning algorithms. Finally, we conclude by discussing how the introduction of $\phi^{4}$ machine learning algorithms opens up a new research avenue, that of developing, computationally and analytically, a framework of machine learning within quantum field theory.

\section{\label{sec:mlphio}The $\phi^4$ scalar field theory as a Markov Random Field}

\begin{figure}[t]
\includegraphics[width=6.5cm]{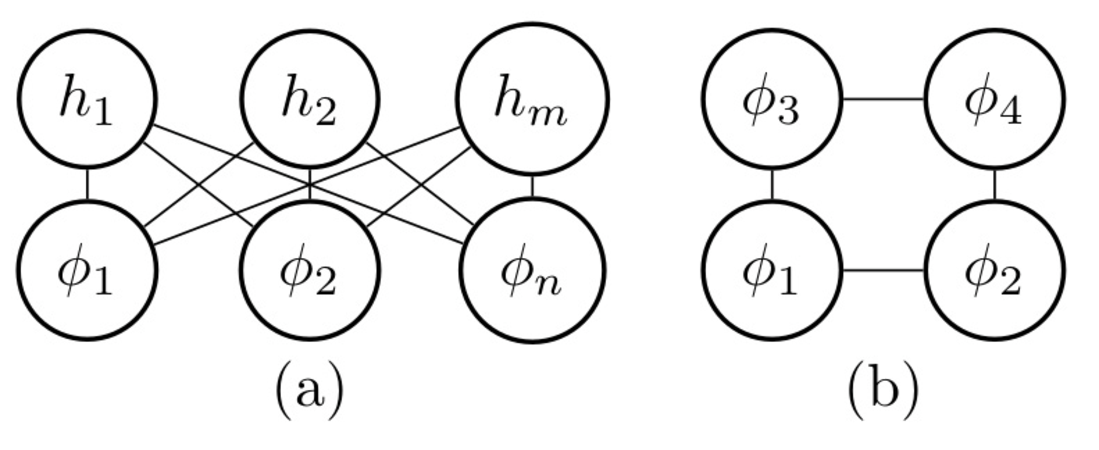}
\caption{\label{fig:graph} (a) A bipartite graph. The maximal cliques correspond to the sites associated with the random variables $\lbrace \phi_{1},h_{1} \rbrace$, $\lbrace \phi_{1},h_{2} \rbrace$, $\lbrace \phi_{1},h_{m} \rbrace$, $\lbrace \phi_{2},h_{1} \rbrace$, $\lbrace \phi_{2},h_{2} \rbrace$, $\lbrace \phi_{2},h_{m} \rbrace$, $\lbrace \phi_{n},h_{1} \rbrace$, $\lbrace \phi_{n},h_{2} \rbrace$, $\lbrace \phi_{n},h_{m} \rbrace$. (b) A square lattice. The maximal cliques correspond to the sites associated with the random variables $\lbrace \phi_{1},\phi_{2} \rbrace$, $\lbrace \phi_{1},\phi_{3}  \rbrace$, $\lbrace \phi_{3},\phi_{4}  \rbrace$ and $\lbrace \phi_{2},\phi_{4}  \rbrace$.  }
\end{figure}

Let $\Lambda$ be a finite set whose points represent the sites of a physical model, and let $\Lambda$ have an additional structure, for instance consider that the spacing between the sites might be known and that the sites are connected. We now consider that the points of $\Lambda$ lie on the vertices of a finite graph $\mathcal{G}=(\Lambda,e)$, where $e$ is the set of edges on $\mathcal{G}$. If $i,j \in \Lambda$ and there exists an edge between $i$ and $j$ then $i$ and $j$ are called neighbours and the set of all neighbours of a considered point $i$ will be denoted by $\mathcal{N}_{i}$.  A clique is a subset of $\Lambda$ where the points are pairwise connected, and a clique is called maximal if no additional point can be included such that the resulting set is still a clique. We will denote a maximal clique as $c$ and the set of all maximal cliques as $C$. For an illustration of the concepts see Fig.~\ref{fig:graph} and for rigorous results see Refs.~\citep{Koller, Preston}.

In addition we associate to each point $i \in \Lambda$ a random variable $\phi_{i, i \in \Lambda}$ and we will call $\phi=\lbrace \phi_{i} \rbrace$ a state or configuration of the system. Given a graph $\mathcal{G}=(\Lambda,e)$, the set of random variables  define a Markov random field if the associated probability distribution $p$ fulfills the local Markov property with respect to $\mathcal{G}$. The local Markov property denotes that a variable $\phi_{i}$ is conditionally independent of all other variables given its neighbors $\mathcal{N}_{i}$, i.e:
\begin{equation}\label{eq:localm}
p(\phi_{i} | (\phi_{j})_{j \in \Lambda -i}) = p(\phi_{i} | (\phi_{j})_{j \in \mathcal{N}_{i}}).
\end{equation}

A probability distribution is then related with the events generated by a Markov random field through the Hammersley-Clifford theorem \citep{Koller}:

\begin{thm}[Hammersley-Clifford.]
A strictly positive distribution $p$ satisfies the local Markov property of an undirected graph $\mathcal{G}$, if and only if $p$ can be represented as a product of strictly positive potential functions $\psi_{c}$ over $\mathcal{G}$, one per maximal clique $c \in C$, i.e.,
\begin{equation}
p(\phi)= \frac{1}{Z} \prod_{c \in C} \psi_{c}(\phi),
\end{equation}
where $Z=\int_{\bm{\phi}} \prod_{c \in C} \psi_{c}(\bm{\phi}) d\bm{\phi}$ is the partition function and $\bm{\phi}$ are all possible states of the system.
\end{thm}

 We will demonstrate that the $\phi^{4}$ scalar field theory satisfies the Hammersley-Clifford theorem and is therefore a Markov random field. The two-dimensional $\phi^{4}$ theory is described by the Euclidean Lagrangian:
\begin{equation}
\mathcal{L}_{E}= \frac{\kappa}{2} (\nabla \phi)^{2} + \frac{\mu_{0}^{2}}{2} \phi^2 + \frac{\lambda}{4} \phi^{4},
\end{equation}
where the action that regularizes the continuum theory on a square lattice is:
\begin{equation}\label{eq:midaction}
S_{E}= -\kappa_{L}\sum_{\langle ij \rangle}\phi_{i} \phi_{j} + \frac{(\mu_{L}^{2}+4\kappa_{L})}{2} \sum_{i} \phi_{i}^{2} +  \frac{\lambda_{L}}{4}  \sum_{i}\phi_{i}^{4}.
\end{equation}

The quantities $\kappa_{L},\mu_{L}^{2},\lambda_{L}$ are dimensionless parameters, one of which is deprecated and can be absorbed by rescaling the fields \citep{Milchev1986}. Nevertheless, consider the set of variables $w=\kappa_{L}$, $a=(\mu_{L}^{2}+4\kappa_{L})/2$, $b=\lambda_{L}/4$ as inhomogeneous and the resulting action as:
\begin{equation}\label{eq:finalaction}
S(\phi ; \theta)= -\sum_{\langle ij \rangle} w_{ij} \phi_{i}\phi_{j} + \sum_{i} a_{i} \phi_{i}^{2} + \sum_{i} b_{i} \phi_{i}^{4},
\end{equation}
where the set of coupling constants is $\theta=\lbrace w_{ij}, a_{i},b_{i} \rbrace$, and the associated Boltzmann probability distribution is:
\begin{equation} \label{eq:probmrf}
p(\phi ; \theta)=\frac{\exp\big[-S(\phi ; \theta)\big]}{\int_{\bm{\phi}}{\exp[-S(\bm{\phi},\theta)]} d\bm{\phi}}.
\end{equation}

 The $\phi^{4}$ scalar field theory is formulated on a graph $\mathcal{G}=(\Lambda,e)$ where $\Lambda$ is the set of lattice sites and $e$ the set of edges or pairwise interactions.  For a square lattice only nearest neighbors define a maximal clique (see Fig.~\ref{fig:graph}). Since we search for arbitrary, strictly positive potential functions $\psi_{c}$ per maximal clique $c \in C$, we can multiply $\psi_{c}$ with strictly positive functions of subsets of $c$ \citep{10.5555/1162264}, i.e. with functions of one-site cliques. We then arrive, after considering the imposed boundary conditions, at a nonunique choice of potential function:
\begin{equation}
\psi_{c} = \exp\bigg[ -w_{ij} \phi_{i} \phi_{j}+ \frac{1}{4} (a_{i} \phi_{i}^{2} +a_{j}\phi_{j}^{2}  +b_{i} \phi_{i}^{4} +b_{j}\phi_{j}^{4})\bigg],
\end{equation}
where $i,j$ are nearest neighbors. As the potential functions $\psi_{c}$ are strictly positive the quantity $\ln \psi_{c} $ can be defined,   and the probability distribution $p(\phi ; \theta)$ can be factorized as:
\begin{equation}
p(\phi ; \theta)=\frac{\exp\big[{\sum_{c \in C} \ln \psi_{c}(\phi)}\big]}{\int_{\bm{\phi}} \exp \big[\sum_{c \in C} \ln\psi_{c}(\bm{\phi})\big]d\bm{\phi}} = \frac{1}{Z} \prod_{c \in C} \psi_{c} (\phi).
\end{equation}

To summarize, the discretized $\phi^{4}$ scalar field theory satisfies the Hammersley-Clifford theorem and the local Markov property and is therefore a Markov random field. To understand intuitively the meaning of the local Markov property, consider the more familiar case satisfied by a Markov chain $P(\phi^{k+1} | \phi^{k},\ldots,\phi^{0})= P(\phi^{k+1}| \phi^{k})$. This property declares that given a certain state $\phi^{k}$ a future state $\phi^{k+1}$ depends only on the current state $\phi^{k}$,  and not on states that preceded it, such as $\phi^{k-1}$. The local Markov property of Eq.~\ref{eq:localm} extends this concept to higher dimensions by giving it a spatial representation via a Markov random field. For the case of the $\phi^{4}$ scalar field theory the variational parameters $\theta$ are the coupling constants $\theta=\lbrace w_{ij}, a_{i},b_{i} \rbrace$.  By considering that the probability $p(\phi ; \theta)$ of the Markov random field depends on the parameters $\theta$  a variety of machine learning tasks can then be completed.

\section{\label{sec:mlphi}Machine Learning with the $\phi^{4}$ scalar field theory}

\subsection{Learning without predefined data}

Consider a target probability distribution $q(\phi)$ of an arbitrary statistical system. An asymmetric measure of the distance between the two probability distributions $p(\phi ; \theta)$ and $q(\phi)$ can be defined, which is called the Kullback-Leibler divergence \citep{Koller}:
\begin{equation} \label{eq:kl}
KL(p || q) = \int_{-\infty}^{\infty} {p(\bm{\phi}; \theta)} \ln \frac{p(\bm{\phi}; \theta)}{q(\bm{\phi})} d\bm{\phi}  \geq 0.
\end{equation}
The Kullback-Leibler divergence is nonnegative and equal to zero when the two probability distributions exactly match one another. We emphasize that the Kullback-Leibler divergence does not satisfy the triangle inequality and it therefore cannot be classified as a proper distance as it is not symmetric. It is the quantity $KL(p || q)+KL(q || p)$ which is a true metric. The Kullback-Leibler divergence will be called an asymmetric distance to retain the intuitive picture that it establishes a measure of the difference between two probability distributions. 

 By searching for an optimal set of coupling constants $\theta=\lbrace w_{ij}, a_{i},b_{i} \rbrace$ we can minimize the Kullback-Leibler divergence so that the probability distribution of the $\phi^4$ scalar field theory $p(\phi ; \theta)$  will converge to the target probability distribution $q(\phi)$. Once minimization is conducted a Markov chain Monte Carlo simulation can be initiated for $p(\phi ; \theta)$ to draw samples that would be representative of the target distribution $q(\phi)$. Let us consider the case where the target probability distribution $q(\phi)$ is that of an arbitrary statistical system with partition function $Z_{\mathcal{A}}$ and it has a Boltzmann form $q(\phi)=\exp[-\mathcal{A}]/Z_{\mathcal{A}}$. Any additional parameter, such as the inverse temperature, is absorbed within the Hamiltonian or lattice action $\mathcal{A}$. By substituting $q(\phi)$ and $p(\phi ; \theta)$ in Eq.~\ref{eq:kl} we arrive at:
\begin{equation} \label{eq:fen}
 -\ln Z_{\mathcal{A}}\leq \langle \mathcal{A} - S \rangle_{p(\phi;\theta)} -\ln Z.
\end{equation}

By considering that the terms $F_{\mathcal{A}} = -\ln Z_{\mathcal{A}} $ and $F = -\ln Z$ are equal to the free energy, the above equation can be equivalently expressed as:
\begin{equation}\label{eq:fen2}
F_{\mathcal{A}} \leq \langle \mathcal{A} - S \rangle_{p(\phi;\theta)} + F \equiv \mathcal{F},
\end{equation}
where $\mathcal{F}$ is the variational free energy. As a result Eq.~\ref{eq:fen2} sets a rigorous upper bound to the calculation of the free energy $F_{\mathcal{A}}$ of the target system and this bound $\mathcal{F}$ is dependent on calculations conducted entirely on the distribution $p(\phi ; \theta)$ of the $\phi^{4}$ Markov random field. This indicates that one can map an arbitrary system to a $\phi^{4}$ scalar field theory by minimizing an asymmetric distance between the probability distributions of the two systems.

A gradient-based approach can then be implemented to minimize the variational free energy $\mathcal{F}$ via its derivatives in terms of the parameters $\theta$:
\begin{equation}
\frac{\partial \mathcal{F}}{\partial \theta_{i}}= \langle \mathcal{A} \rangle \Big\langle \frac{\partial S}{\partial \theta_{i}} \Big\rangle -\Big\langle \mathcal{A} \frac{\partial S}{\partial \theta_{i}} \Big\rangle + \Big\langle S \frac{\partial S}{\partial \theta_{i}} \Big\rangle - \langle S \rangle \Big\langle \frac{\partial S}{\partial \theta_{i}} \Big\rangle ,
\end{equation} 
where all expectation values are calculated under the probability distribution $p(\phi;\theta)$ of the $\phi^{4}$ scalar field theory. Derivations can be found in Appendix~\ref{app:deriv}. The variational parameters are then updated at each epoch $t$ of the minimization process through:
\begin{equation}\label{eq:gas}
\theta^{(t+1)}=\theta^{(t)}-\eta*  \mathcal{L},
\end{equation}
where $\eta$ is the learning rate and $\mathcal{L}=\partial \mathcal{F}/\partial \theta^{(t)}$. After the minimization process we anticipate that $\mathcal{F} \approx F_{A}$ and as a result $p(\phi; \theta) \approx q(\phi)$. 

To illustrate the approach we consider as a target system a $\phi^{4}$ lattice action $\mathcal{A}$ with longer-range interactions and complex-valued coupling constants, defined as: 
\begin{eqnarray}
\mathcal{A}= \sum_{k=1}^{5} g_{k}\mathcal{A}^{(k)}=  g_{1} \sum_{\langle ij \rangle_{nn} } \phi_{i} \phi_{j} + g_{2} \sum_{i} \phi_{i}^{2}  \\+  g_{3} \sum_{i} \phi_{i}^{4}  + g_{4} \sum_{\langle ij \rangle_{nnn} } \phi_{i} \phi_{j} + i g_{5} \sum_{i}  \phi_{i}^{2}.
\end{eqnarray}

The notations $nn$ and $nnn$ denote nearest-neighbor and next-nearest neighbor interactions and the lattice action is complex due to the $ g_{5}\mathcal{A}^{(5)}$ term. The combination of the $g_{2}$ and $g_{5}$ parameters introduces a complex coupling constant in the mass term. The coupling constants have values $g1=g4=-1$, $g_{2}=1.52425$, $g_{3}=0.175$ and $g_{5}=0.15$. The values for $g_{1}$, $g_{2}$ and $g_{3}$ have been chosen near the critical point of the second-order phase transition for the system with a local homogeneous action for which $g_{4}=g_{5}=0$. We will present three applications for lattices of size $L=4$ at each dimension: first, a proof-of-principle demonstration will be conducted to verify that the inhomogeneous action $S$ (see Eq.~\ref{eq:finalaction}) can learn the local lattice action $\mathcal{A}_{\lbrace 3 \rbrace}=\sum_{k=1}^{3} g_{k}\mathcal{A}^{(k)}$. Second, we will discuss that by considering the local lattice action $\mathcal{A}_{\lbrace 3 \rbrace}$ it is impossible to reweight to the full action $\mathcal{A}$ due to insufficient overlap of statistical ensembles, but there exists an inhomogeneous representation of $\mathcal{A}_{\lbrace 3 \rbrace}$ equal to $S$ for which this is possible.  Finally we will demonstrate that $S$ can approximate $\mathcal{A}$ sufficiently to simultaneously extrapolate observables in the parameter space of the complex action $\mathcal{A}$ along the trajectory of a considered coupling constant and we will discuss how to successfully define the allowed reweighting range.

\begin{figure}[t]
\includegraphics[width=8.6cm]{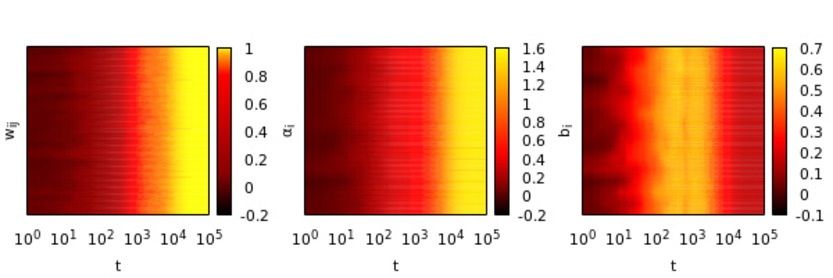}
\caption{\label{fig:heat} Variational parameters $\theta= \lbrace w_{ij},a_{i},b_{i} \rbrace$ versus epochs $t$ on logarithmic scale. The figures depict the evolution of the parameters $\theta$ towards the expected values of the coupling constants in the target homogeneous action.}
\end{figure}

We now initialize the $\phi^{4}$ Markov random field with inhomogeneous coupling constants $\theta$ which are randomly drawn from a Gaussian distribution and consider as a target system in Eq.~\ref{eq:fen} the local lattice action $\mathcal{A}_{\lbrace 3 \rbrace}$. We anticipate that the optimal solution is the one where the inhomogeneous coupling constants $\theta$ of the $\phi^{4}$ Markov random field  will converge to the homogeneous constants $g_{1}$, $g_{2}$ and $g_{3}$ of the target $\phi^{4}$ scalar field theory. Details about the simulations can be found in Appendix~\ref{app:sim}.  The time evolution for the parameters $\theta$  is depicted in Fig~\ref{fig:heat} and details of the training process can be found in Appendix~\ref{app:sim}. After training is conducted the parameters $\theta$ have converged to the homogeneous constants of the target system with precision of order of magnitude of $10^{-8}$ for all cases. It then becomes clear that given sufficient training time the two systems become identical.

\begin{figure}[t]
\includegraphics[width=8.6cm]{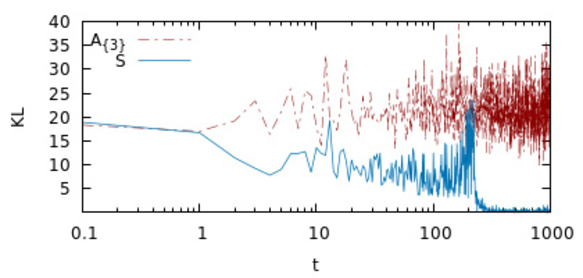}
\caption{\label{fig:kl} Estimated Kullback-Leibler divergence versus epoch $t$ on logarithmic scale. The probability distributions of actions $\mathcal{A}_{\lbrace 3\rbrace }$ and $S$ are compared with the one of $\mathcal{A}_{\lbrace 4\rbrace }$. Only the action $S$ is updated at each epoch based on a finite sample of fixed size. For action $\mathcal{A}_{\lbrace 3\rbrace }$ results are depicted based on a finite sample of equal size to allow for a direct comparison of the two quantities at each epoch $t$. }
\end{figure}

The overlap of statistical ensembles can be quantified through the Kullback-Leibler divergence. We consider the probability distribution $p(\phi;\theta)$, described by the local inhomogeneous action $S$, and we minimize the Kullback-Leibler divergence to approximate the target distribution of action $ \mathcal{A}_{\lbrace 4 \rbrace}$ which is denoted as $q(\phi)$. In addition, we simultaneously estimate the Kullback-Leibler divergence between the distributions of $\mathcal{A}_{\lbrace 3 \rbrace}$ and $ \mathcal{A}_{\lbrace 4 \rbrace}$ to quantify their overlap of statistical ensembles. The results are depicted in Fig.~\ref{fig:kl} where it is evident that the local inhomogeneous action $S$ produces a probability distribution which approximates $ \mathcal{A}_{\lbrace 4 \rbrace}$ exceedingly better than the probability distribution of $\mathcal{A}_{\lbrace 3 \rbrace}$. This tentatively indicates that while $S$ and $\mathcal{A}_{\lbrace 3 \rbrace}$ have the same form of lattice action, the inhomogeneity present in the former allows for the construction of richer representations of probability distributions. As a result,  histogram reweighting~\citep{PhysRevLett.61.2635} from local inhomogeneous actions to regions of parameter space that are inaccessible to the local homogeneous action might be possible.

We proceed to discuss the precise implications of the equivalence between the approximating distribution $p(\phi;\theta)$ of action $S$ and the target distribution $q(\phi)$ of action $ \mathcal{A}_{\lbrace 4 \rbrace}$. The definition of the expectation value $\langle O \rangle_{P}$ of an arbitrary observable $O$ in a system that has some equilibrium occupation probabilities $P$ is:
\begin{equation}\label{eq:estimator}
\langle O \rangle_{P}= \sum_{\bm{\phi}} O_{\bm{\phi}} P({\bm{\phi}}),
\end{equation}
where the sum is over all possible states $\bm{\phi}$ of the system. After the Kullback-Leibler divergence between the distributions $p(\phi;\theta)$ and $q(\phi)$ is minimized $KL \approx 0$ and:
\begin{equation}\label{KLresult}
p(\phi;\theta) \approx q(\phi),
\end{equation}
which instantly implies, based on Eq.\ref{eq:estimator}, that:
\begin{equation}
\langle O \rangle_{p(\phi; \theta)} \approx \langle O \rangle_{q(\phi)}.
\end{equation}

To clarify further, observables, such as the lattice action $ \mathcal{A}_{\lbrace 4 \rbrace}$ should yield approximately equal values when calculated from samples drawn from either distribution $p(\phi ;\theta)$ or $q(\phi)$ even though the two distributions have different actions $S$ and $ \mathcal{A}_{\lbrace 4 \rbrace}$, respectively. To express these ideas in a more formal manner, we now consider the expectation value of an arbitrary observable as obtained during a Monte Carlo simulation (e.g. see Refs~\cite{bachtis2020adding,bachtis2020extending}) in the target system with action $\mathcal{A}_{\lbrace 4 \rbrace}$:
\begin{equation}\label{eq:numestim}
\langle O \rangle_{q(\phi)}= \frac{\sum_{l=1}^{N} \tilde{p}_{{l}}^{-1} O_{{l}} \exp[-\sum_{k=1}^{4}g_{k}\mathcal{A}^{(k)}_{{l}}] }{\sum_{l=1}^{N} \tilde{p}_{{l}}^{-1} \exp[-\sum_{k=1}^{4}g_{k}\mathcal{A}^{(k)}_{{l}}]},
\end{equation}
where $\tilde{p}$ are the probabilities used to sample from the equilibrium distribution and $N$ the number of samples that we have obtained during the Monte Carlo simulation. There are two fundamentally different ways to proceed in calculating the expectation value of the above equation by relying instead on the approximating probability distribution $p(\phi;\theta)$. 

\begin{figure}[t]
\includegraphics[width=8.6cm]{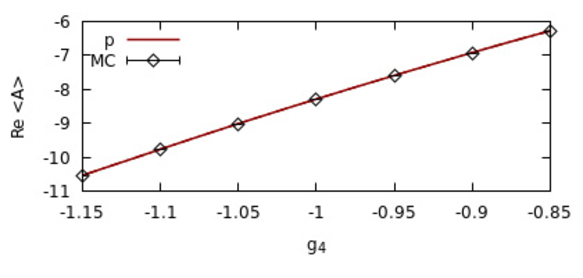}
\caption{\label{fig:re1} Real part of the complex lattice action $\mathcal{A}$ versus coupling constant $g_{4}$. The results are obtained by reweighting from the Markov random field distribution $p$ to the distribution of the complex action $\mathcal{A}$. The statistical errors are comparable with the width of the line.   The results are compared with Monte Carlo (MC) and reweighting from the distribution of the real action $\mathcal{A}_{\lbrace 4\rbrace }$ to $\mathcal{A}$.}
\end{figure}

The first is to draw a subset of samples from $p(\phi;\theta)$ and then conjecture, based on Eq.~\ref{KLresult}, that these $N$ samples have been produced instead by the distribution $q(\phi)$. This would have been equivalent to considering $\tilde{p}=q(\phi)$ in Eq.~\ref{eq:numestim} but a systematic error would be introduced based on the accuracy in which the probability distribution $p(\phi;\theta)$ approximates $q(\phi)$. The second approach again relies on drawing a subset of samples from the distribution $p(\phi;\theta)$, but this time we will consider that $p(\phi;\theta) \neq q(\phi)$ and that the samples have been produced directly from $p(\phi;\theta)$ of Eq.~\ref{eq:probmrf} with action $S$. This is equivalent to conducting a reweighting step so that the probability distribution $p(\phi;\theta)$ will become equal to the distribution $q(\phi)$ under the condition that there exists a sufficient overlap of ensembles between the two distributions. We anticipate that this reweighting step is possible to achieve due to the minimization of the Kullback-Leibler divergence between the two distributions $p(\phi;\theta)$ to $q(\phi)$ and their approximate equivalence.

We will follow the second approach and implement a reweighting technique, details of which can be found in Appendix~\ref{app:reweig}, to simultaneously extrapolate observables in the parameter space of the full action $\mathcal{A}$ which includes complex couplings and longer-range interactions:
\begin{equation}\label{eq:rewfull}
\langle O \rangle= \frac{\sum_{l=1}^{N} O_{{l}} \exp[S_{{l}}-g_{j}'\mathcal{A}_{{l}}^{(j)}- \sum_{k=1,k \neq j}^{5}g_{k}\mathcal{A}^{(k)}_{{l}}]}{\sum_{l=1}^{N}  \exp[S_{{l}}-g_{j}'\mathcal{A}_{{l}}^{(j)}- \sum_{k=1,k \neq j}^{5}g_{k}\mathcal{A}_{{l}}^{(k)}]}.
\end{equation}

The equation above can be interpreted as two distinct simultaneous reweighting steps. First the probability distribution $p(\phi; \theta)$ of the $\phi^{4}$ Markov random field with action $S$ is reweighted to the distribution $q(\phi)$ with action $ \mathcal{A}_{\lbrace 4 \rbrace}$ but with a shifted coupling constant $g_j'$. This acts as a correction step to ensure that the proper distribution is reached from $p(\phi;\theta)$ and it additionally allows an extrapolation along the direction of the parameter space described by coupling $g_j'$. Second there is a reweighting step to reach the distribution described by the complex lattice action $\mathcal{A}$, which includes the imaginary part $g_{5} \mathcal{A}^{(5)}$. Any arbitrary observable can be reweighted in parameter space, such as machine learning derived observables~\citep{bachtis2020extending}, and Hamiltonian-agnostic reweighting~\citep{bachtis2020adding} could additionally be explored.

\begin{figure}[t]
\includegraphics[width=8.6cm]{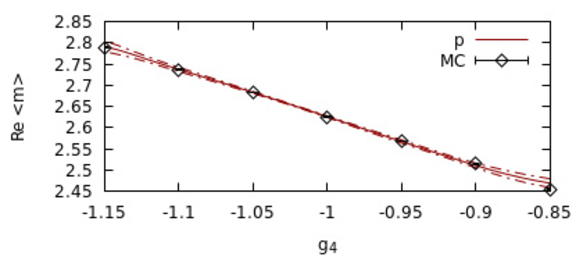}
\caption{\label{fig:re2}   Real part of the magnetization $m$ versus coupling constant $g_{4}$. The results are obtained by reweighting from the Markov random field distribution $p$ to the target distribution of the complex action $\mathcal{A}$. The associated statistical errors are depicted by the dashed lines. The results are compared with Monte Carlo (MC) and reweighting from the distribution of the real action $\mathcal{A}_{\lbrace 4\rbrace }$ to $\mathcal{A}$.}
\end{figure}

We consider that $j=4$ and we extrapolate observables along the trajectory of the $g_{4}'$  coupling constant for a continuous range of values $g_{4}' \in[-0.85,-1.15]$. We recall that the $\phi^{4}$ Markov random field was trained to approximate the action $\mathcal{A}_{\lbrace 4 \rbrace}$ where $g_{4}=-1$.  Results for the magnetization and the internal energy, obtained with reweighting from the probability distribution $p(\phi;\theta)$ to the full action $\mathcal{A}$ are depicted in Figs.~\ref{fig:re1} and~\ref{fig:re2}. The results are compared with Monte Carlo simulations conducted on action $\mathcal{A}_{\lbrace 4 \rbrace}$ which are combined with reweighting to the full complex distribution to allow for a comparison with the ones from $p(\phi;\theta)$. It is evident that the results depicted agree within statistical errors with the Monte Carlo extrapolations. Details about the statistical error analysis can be found in Appendix~\ref{app:err}.

When reweighting is implemented to extrapolate to the probability distribution of a complex action or as a correction step in the case of an approximating distribution the question of how to strictly define the reweighting range emerges. This can be achieved, formally, through the calculation of weight functions which are dependent on the underlying histograms. Specifically, we consider as an example in Eq.~\ref{eq:rewfull} the expectation value of the action $S$.  In addition, instead of expressing Eq.~\ref{eq:rewfull} as a sum over each action $S_{l}$ calculated on a configuration $\phi$ we instead reformulate it in terms of each uniquely sampled action $S$ in the Monte Carlo data set after the construction of histograms. The expectation value is then:
\begin{equation}
\langle S \rangle = \sum_{S} S \mathcal{W}(S),
\end{equation}
where the sum is over uniquely sampled actions $S$ and $\mathcal{W}(S)$ is a weight function which is equal to:

\begin{widetext}

\begin{equation}
\mathcal{W}(S)=  \frac{\sum_{\Re [\mathcal{A'}], \Im [\mathcal{A'}]}  h(S,\Re [\mathcal{A'}], \Im [\mathcal{A'}])  \exp[S-\Re [\mathcal{A'}]- i \Im [\mathcal{A'}]]}{\sum_{S,\Re [\mathcal{A'}], \Im [\mathcal{A'}]} h(S,\Re [\mathcal{A'}], \Im [\mathcal{A'}])  \exp[S-\Re [\mathcal{A'}]-i\Im [\mathcal{A'}]]},
\end{equation}
\end{widetext}

\begin{figure}[b]
\includegraphics[width=8.6cm]{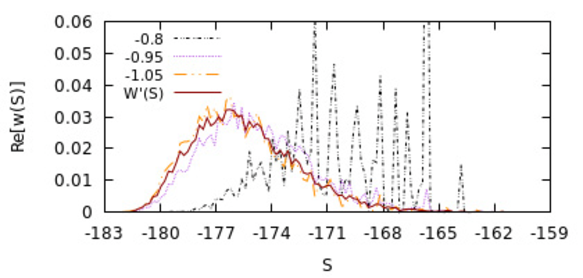}
\caption{\label{fig:hist} Real part of the weight function $\mathcal{W}(S)$ versus lattice action $S$ for considered coupling constants $g_{4}' \in [-1.05,-0.8]$. The results are obtained by reweighting from the local inhomogeneous action $S$ to the complex action $\mathcal{A}$ which includes longer-range interactions. }
\end{figure}

where $\mathcal{A'}=g_{j}'\mathcal{A}^{(j)}+ \sum_{k=1,k \neq j}^{5}g_{k}\mathcal{A}^{(k)}$. The quantity $h(S,\Re [\mathcal{A'}], \Im [\mathcal{A'}])$ is a multi-dimensional histogram of the inhomogeneous action $S$ as well as each action term in which we are interested to extrapolate towards during reweighting. Reweighting can be achieved either by including novel terms in the action or by shifting its corresponding coupling constant if the term already exists. Of particular interest is also the quantity $\mathcal{W}'(S)$ where the exponentials are chosen equal to one and which is proportional to the actual histograms of the action in the corresponding Monte Carlo data set. This quantity can additionally serve as an indication of the reweighting range.

\begin{figure}[b]
\includegraphics[width=8.6cm]{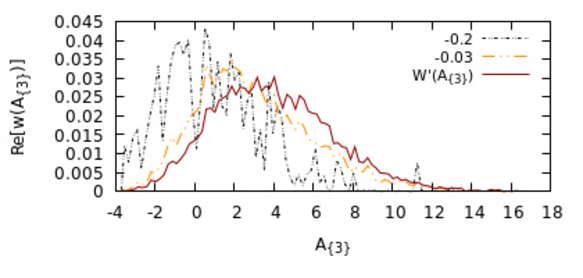}
\caption{\label{fig:hist2} Real part of the weight function $\mathcal{W}(\mathcal{A}_{\lbrace 3 \rbrace})$ versus lattice action $\mathcal{A}_{\lbrace 3 \rbrace}$ for considered coupling constants $g_{4}'$. The results are obtained by reweighting from action $\mathcal{A}_{\lbrace 3 \rbrace}$ to the complex action $\mathcal{A}$ which includes longer-range interactions. }
\end{figure}

We proceed to calculate the weight functions $\mathcal{W}(S)$ for each uniquely sampled action $S$ in a considered extrapolation range. The results are depicted in Fig.~\ref{fig:hist} where an overlap between distinct weight functions that are adjacent in parameter space to the coupling constant $g_{4}=-1$ is observed.  We recall that reweighting extrapolations are accurate only when the method successfully predicts the form of histograms at the extrapolated point in parameter space based on the histograms present at the initial data set. When the coupling constant is $g_{4}'=-0.8$ major inconsistencies can be noticed. This indicates that reweighting extrapolations to $g_{4}'=-0.8$ would be inaccurate as the form of the weight functions cannot be successfully predicted. 

We emphasize that reweighting from the local homogeneous action $\mathcal{A}_{\lbrace 3 \rbrace}$ to the full action $\mathcal{A}$ is not possible. The inclusion of an imaginary term and a longer range interaction does not produce a sufficient overlap of ensembles. Results are depicted in Fig.~\ref{fig:hist2}. We recall that the local homogeneous action $\mathcal{A}_{\lbrace 3 \rbrace}$ has coupling constant $g_{4}=0$ and the target distribution of action $\mathcal{A}_{\lbrace 4 \rbrace}$  includes a term with coupling constant $g_{4}'=-1.0$. It is clear that the values of the lattice action lie at an entirely different scale and inconsistencies begin to emerge when $g_{4}'=-0.2$. Reweighting to the full action is then impossible from the probability distribution of action $\mathcal{A}_{\lbrace 3 \rbrace}$. However, the local inhomogeneous action $S$ is able to achieve reweighting to the full distribution of the action $\mathcal{A}$. Consequently the opportunity to map improved lattice actions, which include longer-range interactions, to local inhomogeneous actions is a prospect that is open to explore. This can be achieved by minimizing the asymmetric distance between their associated probability distributions.

\subsection{Learning with predefined data}

\begin{figure}[t]
\includegraphics[width=7.6cm]{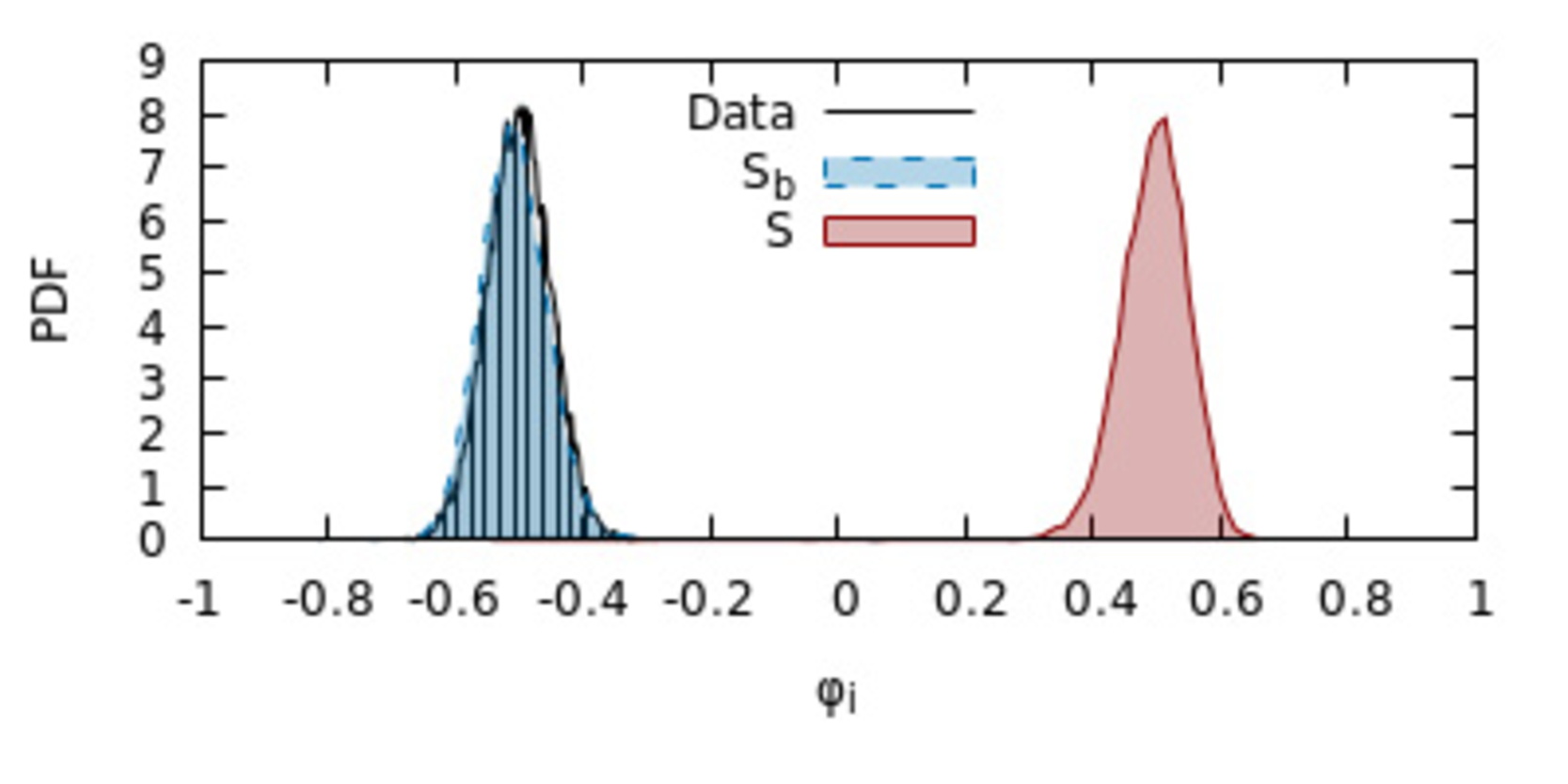}
\caption{\label{fig:2} Probability density function versus lattice value $\phi_{i}$ for a Euclidean action $S$ that is $Z_{2}$ invariant and $S_{b}$ which includes a local symmetry-breaking term.}
\end{figure}

The preceding results do not require any predefined data to be used as input within the training process since configurations were obtained during the gradient-based approach.
 However, there exist cases where one has already obtained a set of available data, which could comprise configurations of a system, experimental data, or a set of images, and whose probability distribution is of unknown form. The obtained data set then explicitly encodes an empirical probability distribution $q(\phi)$ that is a representation of the complete probability distribution of the system. The empirical distribution $q(\phi)$ can still be learned by minimizing instead the opposite divergence:
\begin{equation} \label{eq:klopp}
KL(q || p) = \int_{-\infty}^{\infty} {q(\bm{\phi})} \ln \frac{q(\bm{\phi})}{p(\bm{\phi}; \theta)} d\bm{\phi}  \geq 0.
\end{equation}

By expanding the above equation we arrive at:
\begin{equation}\label{eq:klimg}
KL(q || p)= \langle \ln q(\phi) \rangle_{q(\phi)}- \langle \ln p(\phi;\theta) \rangle_{q(\phi)}.
\end{equation} 

The first right-hand term is constant and the minimization of $KL(q || p)$ is therefore equivalent to the maximization of the second right-hand term under the training data: 
\begin{equation}
\frac{\partial \ln p(\phi ; \theta)}{\partial \theta} = \Big\langle \frac{\partial S}{\partial \theta} \Big\rangle_{p(\phi ; \theta)} -  \frac{\partial S}{\partial \theta}.
\end{equation}

The variational parameters are now updated according to Eq.~\ref{eq:gas} where $\mathcal{L}=-\partial \ln p(\phi ; \theta^{(t)})/{\partial \theta^{(t)}}$.

To illustrate the concepts we now create a data set from a Gaussian distribution with $\mu=-0.5$ and $\sigma=0.05$ which encodes an empirical distribution $q(\phi)$. The information about the form of $q(\phi)$ will not be introduced in Eq.~\ref{eq:klopp} because the training will instead be conducted on the obtained data. To clarify further, the same approach can be established for any obtained data set, without the need to even infer the underlying form of the distribution. After successful training, Markov chain Monte Carlo simulations can be implemented based on the distribution $p(\phi;\theta)$ of the $\phi^{4}$ Markov random field to draw samples that would be representative of the unknown target distribution $q(\phi)$. Additional details can be found in Appendix~\ref{app:deriv}.

 We anticipate, due to the invariance under the $Z_{2}$ symmetry in the lattice action $S$, that the symmetric distribution with $\mu=0.5$ might be additionally reproduced. If this feature is not desirable then a local symmetry-breaking term of the form $\sum_{i} r_{i} \phi_{i}$ can be included in the action $S$ to favor configurations that will explicitly reproduce $q(\phi)$. The Hammersley-Clifford theorem is still satisfied and results for the symmetric action $S$ and the action $S_{b}$ which includes a symmetry-breaking term are depicted in Fig.~\ref{fig:2}. We observe for the symmetric case that while the algorithm has been trained on one of the probable solutions it is able to produce additional solutions that are invariant under the inherent symmetry, whereas this feature has been eliminated for the broken-symmetry case where the probability distribution $q(\phi)$ is explicitly reproduced.

\begin{figure}[t]
\includegraphics[width=7.2cm]{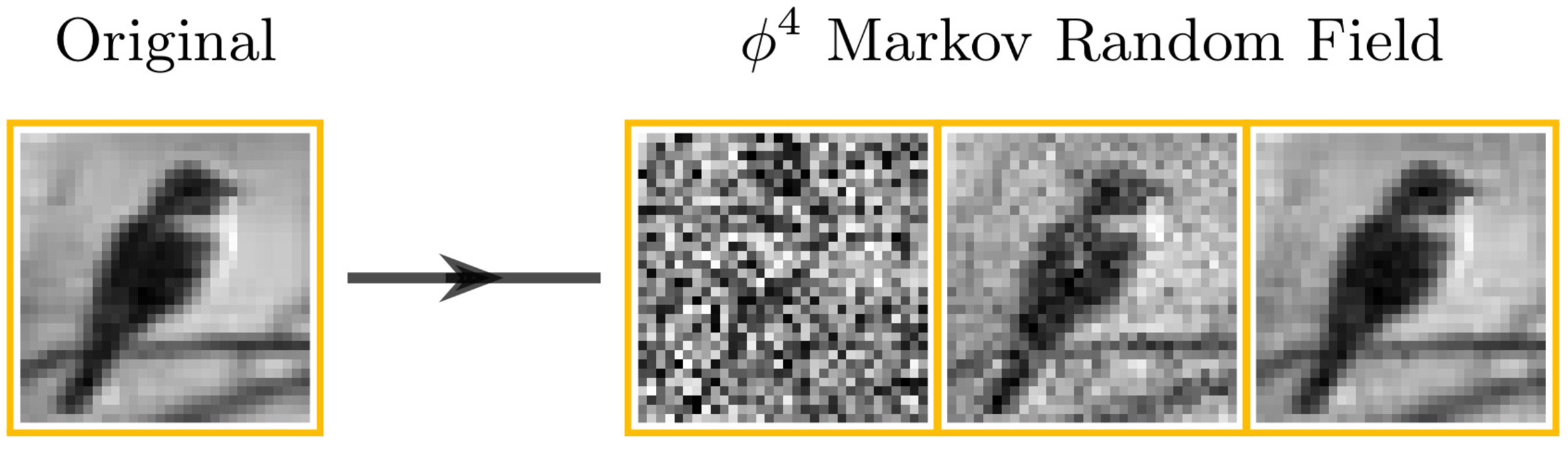}
\caption{\label{fig:bird} Original image and equilibration of the Markov random field after $1$, $10$ and $50$ steps.}
\end{figure}
Markov random fields are widely applied to problems in computer vision, image segmentation and compression, as well as image analysis~\citep{10.5555/2024611}. Every problem that is formulated as an energy or lattice action minimization problem can be solved by implementing Markov random fields. Since the $\phi^{4}$ scalar field theory satisfies the Hammersley-Clifford theorem and is therefore a Markov random field it can be implemented to complete such tasks. We therefore consider as $q(\phi)$ in Eq.~\ref{eq:klimg} the configuration of an image from the CIFAR-10 data set~\citep{Krizhevsky2009LearningML}, which we will map to the action of the inhomogeneous $\phi^{4}$ theory of Eq.~\ref{eq:finalaction}. In essence, we search for the optimal values of the coupling constants, which describe the local interactions in the $\phi^{4}$ scalar field theory, that can reproduce the considered image as a configuration in the equilibrium distribution of the system. We emphasize that the coupling constants $w_{ij}$ relate two adjacent lattice sites and are therefore of utmost importance in uncovering the spatial structure of the image. In Fig.~\ref{fig:bird}, results are depicted after training the $\phi^{4}$ theory. We observe that by initializing a Markov chain the configurations of the equilibrium distribution converge to an accurate representation of the original image. 

\section{$\phi^{4}$ Neural Networks}
When the aim of the machine learning task is to study intricate probability distributions, deep learning algorithms that include multiple layers in the neural network architecture can be implemented. These layers progressively transform data to arrive at increasingly abstract representations, allowing for increased expressivity and representational capacity in the model. Such cases of deep learning algorithms can be constructed from the dynamics of the $\phi^{4}$ scalar field theory. 

We consider that part of the random variables $\phi_{i}$ on the lattice sites are visible and correspond to a set of observations and the remaining are hidden variables $h_{j}$, which capture dependencies on a set of training data, given as input to $\phi_{i}$. In addition, to make the connection with the computer science literature we consider a bipartite graph which imposes the restriction that interactions are exclusively between the $\phi$ and the $h$ variables (see Fig.~\ref{fig:graph}). We therefore recast the $\phi^{4}$ neural network as a variant of a restricted Boltzmann machine (RBM) \citep{10.5555/104279.104290,ACKLEY1985147,FISCHER201425,Hinton2012}, which is able to model continuous data. Alternative parametrizations of the graph structure are open to explore.  A joint probability distribution $p(\phi,h ; \theta)$ is then defined, based on a lattice action $S (\phi,h ; \theta)$: 
\begin{align}\label{eq:many}
S (\phi,h ; \theta) =  -\sum_{i,j} w_{ij} \phi_{i}h_{j}  + \sum_{i} r_{i} \phi_{i}  + \sum_{i} a_{i} \phi_{i}^{2} \\+ \sum_{i} b_{i} \phi_{i}^{4}  + \sum_{j} s_{j} h_{j} + \sum_{j} m_{j} h_{j}^{2} + \sum_{j} n_{j} h_{j}^{4},
\end{align}
which also gives rise to a new expression, based on Eq.~\ref{eq:klopp}, for the derivative of the log-likelihood $\ln p(\phi,\theta)$:
\begin{equation}
\frac{\partial  \ln p(\phi; \theta)}{\partial \theta}  =\Big\langle \frac{\partial S}{\partial \theta}  \Big\rangle_{p(\phi, h ; \theta)}-\Big\langle \frac{\partial S }{\partial \theta} \Big\rangle_{p(h | \phi ; \theta)} ,
\end{equation}
where the set of variational parameters is now $\theta=\lbrace w_{ij},r_{i},a_{i},b_{i},s_{j},m_{j},n_{j}\rbrace$.  The conditional distributions of the visible and the hidden variables are $p(\phi | h ; \theta)= \prod_{i} p(\phi_{i} | h)$ and $p(h | \phi ; \theta)= \prod_{j} p(h_{j} | \phi)$. Derivations can be found in Appendix~\ref{app:deriv}.

By considering certain values of parameters in the $\phi^{4}$ neural network of Eq.~\ref{eq:many} one can arrive at other neural network architectures, all of which are special cases of a $\phi^{4}$ Markov random field. For instance by choosing $b_i=n_j=0$ one obtains a Gaussian-Gaussian RBM \citep{FISCHER201425,10.5555/104279.104290}. If $b_i=n_j=m_j=0$ and $h_{j} \in \lbrace-1,1\rbrace$ then the architecture is a Gaussian-Bernoulli RBM\citep{FISCHER201425,10.5555/104279.104290}. Of particular interest could be the choice of $m_{j}=n_{j}=0$ and $h_{j} \in \lbrace-1,1 \rbrace$ which would reduce to a $\phi^{4}$-Bernoulli RBM, a case with a nonlinear sigmoid function that, to our knowledge, has not been studied before. We emphasize that the $\phi^{4}$ Bernoulli RBM is anticipated to have substantial representational capacity due to the presence of the nonlinear sigmoid function in the hidden layer \citep{10.5555/2976456.2976476}.

 It is a well-known fact that the $\phi^{4}$ scalar field theory of Eq.~\ref{eq:midaction}, a model with continuous degrees of freedom, reduces to an Ising model under the limit $\kappa_{L}$ fixed, $\lambda_{L} \rightarrow \infty$ and $\mu_{L}^{2} \rightarrow - \infty$ \citep{Milchev1986}. The $\phi^{4}$-Bernoulli RBM can then be interpreted as a $\phi^{4}$ neural network where certain lattice sites have reached the Ising limit, allowing for novel physical insights. It is important to recall that, with the inclusion of two hidden layers, deep variants of restricted Boltzmann machines are universal approximators of probability distributions \citep{pmlr-v28-krause13}.

\begin{figure}[t]
\includegraphics[width=8.6cm]{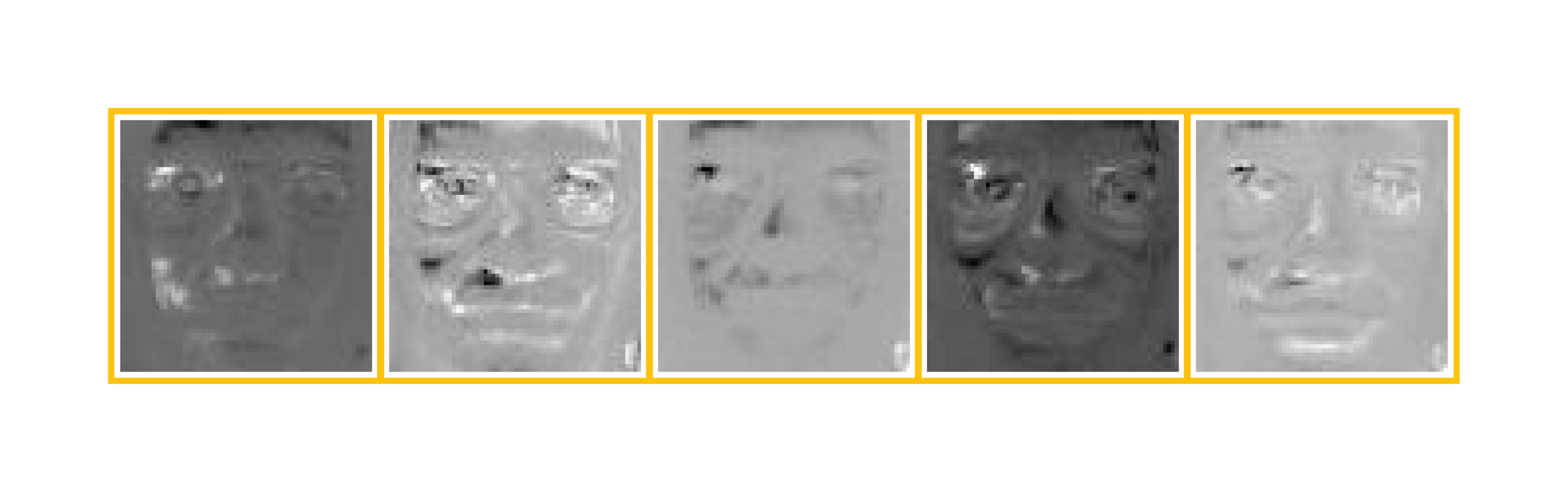}
\caption{\label{fig:faces} Example features learned in the hidden layer of the $\phi^{4}$ neural network.}
\end{figure}

To demonstrate the applicability of the $\phi^{4}$ neural network of Eq.~\ref{eq:many}, we train it on the first forty examples of the Olivetti faces data set \footnote{This data set contains a set of face images taken between April 1992 and April 1994 at AT\&T Laboratories Cambridge} using $4096$ visible units and $32$ hidden units to observe if meaningful features are learned. A subset of the learned features, i.e. the coupling constants $w_{ij}$ for a fixed $j$, are depicted in Fig.~\ref{fig:faces}. We observe that the neural network has learned hidden features which comprise abstract face shapes and characteristics.  The hidden units can then serve as input to a new $\phi^{4}$ neural network to progressively extract abstract features in data \citep{Hinton504}. 

\section{Conclusions}

In this manuscript we derived machine learning algorithms from discretized Euclidean field theories. Specifically we demonstrated that the $\phi^{4}$ scalar field theory on a square lattice satisfies the Hammersley-Clifford theorem and is therefore a Markov random field that can be used for inference and learning. By recasting the $\phi^{4}$ theory within a mathematically rigorous framework a variety of theorems, as well as training algorithms, are available and an overview can be found in Ref.~\citep{Koller}. As the resulting algorithm has inhomogeneous coupling constants it can additionally be investigated from the perspective of spin glasses and of quenched disorder \citep{HANDS1988597,Mezard1987,PhysRevLett.121.071601,PhysRevD.98.045012}, and enhanced sampling can be obtained based on computational techniques from statistical mechanics \citep{Marinari_1992,PhysRevLett.71.211}, or model-specific algorithms \citep{PhysRevLett.62.1087,PhysRevD.58.076003}.  

The Kullback-Leibler divergence can be utilized to quantify the overlap of statistical ensembles between probability distributions. Specifically, we demonstrated that the $\phi^{4}$ scalar field theory with inhomogeneous coupling constants is able to absorb longer-range interactions and observables can be reweighted to the parameter space of complex actions using the approximating probability distribution. The results have been obtained on small lattice volumes. On larger systems more demanding simulations would be needed to achieve the required precision within the permitted reweighting range.  The prospect of constructing improved lattice actions~\citep{BIETENHOLZ1997921,BIETENHOLZ1998114} based on local inhomogeneous representations is open to explore.

 In principle any arbitrary system can be mapped to a $\phi^{4}$ scalar field theory with inhomogeneous coupling constants by minimizing an asymmetric distance of their probability distributions based on Eq.~\ref{eq:fen}.  The concepts are therefore anticipated to be generally applicable to systems within condensed matter physics, lattice field theories and statistical mechanics.   To enhance the accuracy a variant of a neural network architecture can be implemented which is proven to be a universal approximator of a probability distribution \citep{pmlr-v28-krause13}. In the manuscript such variants have been presented as special cases of a $\phi^{4}$ neural network. 

The resulting $\phi^{4}$ machine learning algorithm of Sections~\ref{sec:mlphio} and ~\ref{sec:mlphi} retains the topology of the lattice structure and the boundary conditions, but differs from the conventional $\phi^{4}$ scalar field theory due to the inhomogeneous coupling constants. To employ the tools of quantum field theory a framework involving the replica method is required, but the theories can still be formulated in terms of the functional integral with an additional averaging over the space of couplings \citep{Jain2016}.  It is noted that in our formulation the couplings are inhomogeneous but not random as they are determined during the minimization process.

 We emphasize that prior arguments considering the Hammersley-Clifford theorem hold for arbitrary dimensions and one could therefore construct a $d$-dimensional Markov random field to initiate analytical or computational investigations. The factorization of a lattice action in terms of products of potential functions, a step that is required to recast a system as a Markov random field, depends on the topology of the graph structure and different topologies yield different maximal cliques.   An equivalence between local, pairwise and global Markov properties of a graph structure  can also be rigorously proven \citep{Koller}.  Through the construction of quantum fields in Minkowski space from Markov fields in Euclidean space \citep{NELSON197397}, a new research avenue is envisaged, namely that of developing a computational and mathematical framework of machine learning within quantum field theory.

\section{\label{sec:level5}Acknowledgements}
The authors received funding from the European Research Council (ERC) under the European Union's Horizon 2020 research and innovation programme under grant agreement No 813942. The work of GA and BL has been supported in part by the UKRI Science and Technology Facilities Council (STFC) Consolidated Grant ST/P00055X/1. The work of BL is further supported in part by the Royal Society Wolfson Research Merit Award WM170010 and by the Leverhulme Foundation Research Fellowship RF-2020-461\textbackslash 9. Numerical simulations have been performed on the Swansea SUNBIRD system. This  system is part of the Supercomputing Wales project, which is part-funded by the European Regional Development Fund (ERDF) via Welsh Government. We thank COST Action CA15213 THOR for support.

\appendix

\section{Derivations \label{app:deriv}}

\subsection{$\phi^{4}$ Markov random field}

The Kullback-Leibler divergence, which is repeated here for convenience, defines an asymmetric measure of the distance between the distribution of the machine learning algorithm $p(\phi ; \theta)$ and an unknown target distribution $q(\phi)$:
\begin{equation} \label{eq:kull3}
KL(p || q) =   \int_{-\infty}^{\infty} p(\bm{\phi};\theta) \ln \frac{p(\bm{\phi};\theta)}{q(\bm{\phi})} d\bm{\phi}  \geq 0.
\end{equation}

By expanding the above equation we arrive at:
\begin{equation}
\langle \ln p(\phi;\theta) \rangle_{p(\phi;\theta)}-\langle  \ln q(\phi) \rangle_{p(\phi;\theta)} \geq 0,
\end{equation}
where $\langle \rangle_{p(\phi;\theta)}$ denotes the expectation value under the probability distribution $p(\phi;\theta)$. If the two probability distributions are substituted to be of Boltzmann form, $p(\phi;\theta)=\exp [-S]/Z$, $q(\phi)=\exp[-\mathcal{A}]/Z_{\mathcal{A}}$, we arrive at:
\begin{equation}
-\langle   \ln Z_{\mathcal{A}} \rangle_{p(\phi;\theta)} \leq  \langle \mathcal{A}-S \rangle_{p(\phi;\theta)} - \langle \ln Z \rangle_{p(\phi;\theta)}.
\end{equation}

The terms $\langle \ln Z \rangle_{p(\phi;\theta)} $ are constant in terms of expectation values and we therefore obtain:

\begin{equation}
-   \ln Z_{\mathcal{A}}  \leq  \langle \mathcal{A}-S \rangle_{p(\phi;\theta)} -  \ln Z .
\end{equation}

By denoting the right-hand part as $\mathcal{F}$, the derivative in terms of a variational parameter $\theta_{i}$ is equal to:
\begin{widetext}
\begin{equation}
\frac{\partial \mathcal{F}}{\partial \theta_{i}}= \frac{\partial \langle \mathcal{A} \rangle_{p(\phi; \theta)}}{\partial  \theta_{i}}- \frac{\partial \langle S \rangle_{p(\phi;\theta)}}{\partial  \theta_{i}} - \frac{\partial  (-\ln Z)}{\partial  \theta_{i}},
\end{equation}
where each term is calculated as:

\begin{align}
\frac{\partial \langle \mathcal{A} \rangle_{p(\phi; \theta)}}{\partial  \theta_{i}} &= \frac{\partial}{\partial \theta_{i}}\Bigg[\frac{\int_{\bm{\phi}} \mathcal{A}(\bm{\phi}) \exp [- S(\bm{\phi};\theta)] d\bm{\phi}}{\int_{\bm{\phi}} \exp [- S(\bm{\phi};\theta)]d\bm{\phi}}   \Bigg] = -\Big\langle \mathcal{A}  \frac{\partial  S}{\partial \theta_{i}}\Big\rangle_{p(\phi;\theta)} + \langle \mathcal{A} \rangle_{p(\phi;\theta)} \Big\langle \frac{\partial  S}{\partial \theta_{i}} \Big\rangle_{p(\phi;\theta)},
\end{align}
\begin{align}
\frac{\partial \langle S \rangle_{p(\phi; \theta)}}{\partial  \theta_{i}} &= \frac{\partial}{\partial \theta_{i}}\Bigg[\frac{\int_{\bm{\phi}} S(\bm{\phi} ;\theta) \exp [- S(\bm{\phi};\theta)] d\bm{\phi}}{\int_{\bm{\phi}} \exp [- S(\bm{\phi};\theta)] d\bm{\phi}}   \Bigg] =\Big\langle  \frac{\partial  S}{\partial \theta_{i}}\Big\rangle_{p(\phi;\theta)}  -\Big\langle S  \frac{\partial  S}{\partial \theta_{i}}\Big\rangle_{p(\phi;\theta)} + \langle S \rangle_{p(\phi;\theta)} \Big\langle \frac{\partial  S}{\partial \theta_{i}} \Big\rangle_{p(\phi;\theta)},
\end{align}
\begin{align}
\frac{\partial (- \ln Z)}{\partial  \theta_{i}} & =- \frac{\int_{\bm{\phi}} \frac{\partial}{\partial \theta_{i}}(-S(\bm{\phi};\theta))  \exp[-S(\bm{\phi};\theta) ]d\bm{\phi}}{\int_{\bm{\phi}} \exp[-S(\bm{\phi};\theta)] d\bm{\phi}}= \Big\langle \frac{\partial S}{\partial \theta_{i}} \Big\rangle_{p(\phi;\theta)},
\end{align}

By substituting we arrive at:
\begin{equation}
\frac{\partial \mathcal{F}}{ \partial \theta_{i}}= -\Big\langle \mathcal{A}  \frac{\partial  S}{\partial \theta_{i}}\Big\rangle + \langle \mathcal{A} \rangle \Big\langle \frac{\partial  S}{\partial \theta_{i}} \Big\rangle- \cancel{\Big\langle  \frac{\partial  S}{\partial \theta_{i}}\Big\rangle} +\Big\langle S  \frac{\partial  S}{\partial \theta_{i}}\Big\rangle - \langle S \rangle \Big\langle \frac{\partial  S}{\partial \theta_{i}} \Big \rangle + \cancel{\Big\langle \frac{\partial S}{\partial \theta_{i}} \Big\rangle}.
\end{equation}

A gradient based approach can be implemented based on the above equation to learn a target known probability distribution. 

In the opposite direction if a set of data is available for which the probability distribution is unknown the alternative Kullback-Leibler divergence can be considered:

\begin{equation} \label{eq:kull2}
KL(q || p) =   \int_{-\infty}^{\infty} q(\bm{\phi}) \ln \frac{q(\bm{\phi})}{p(\bm{\phi}; \theta)} d\bm{\phi} .
\end{equation}

By expanding the right-hand side we arrive at the expression:
\begin{equation} \label{eq:kullback}
KL(q || p) =  \langle \ln q(\phi) \rangle_{q(\phi)} - \langle  \ln p(\phi ; \theta)  \rangle_{q(\phi)}.
\end{equation}

 Minimizing the Kullback-Leibler divergence is equivalent to the maximization of the term $\langle \ln p(\phi ; \theta) \rangle_{q(\phi)} $, which is:
\begin{equation}
\langle \ln p(\phi ; \theta) \rangle_{q(\phi)}= \frac{1}{N} \sum_{x} \ln p(\phi^{(x)} ; \theta) ,
\end{equation}
where $x$ is a training example and $N$ the number of training data. For the case of the Markov random field the derivative of the log-likelihood is:

\begin{align*}
\frac{\partial \ln p(\phi ; \theta) }{ \partial \theta} &= \frac{\partial }{\partial \theta} \Bigg[ \ln \frac{\exp [-S(\phi ; \theta)]}{\int_{\bm{\phi}} \exp [-S(\bm{\phi} ; \theta)] d\bm{\phi} } \Bigg] = \frac{\partial }{\partial \theta} \bigg[ \ln \exp [-S(\phi ; \theta)] - \ln \int_{\bm{\phi}} \exp [-S(\bm{\phi} ; \theta)] d\bm{\phi} \bigg] \\ &= \frac{\partial }{\partial \theta} (-S(\phi ; \theta)) - \frac{\int_{\bm{\phi}}  \frac{\partial }{\partial \theta} (-S(\bm{\phi} ; \theta)) \exp  [-S(\bm{\phi} ; \theta)] d\bm{\phi}}{\int_{\bm{\phi}}   \exp  [-S(\bm{\phi} ; \theta)] d\bm{\phi}}= \frac{\partial }{\partial \theta} (-S(\phi ; \theta)) - \int_{\bm{\phi}} p(\bm{\phi} ; \theta) \frac{\partial (-S(\bm{\phi} ; \theta)) }{\partial \theta} d\bm{\phi} \\ & =  \frac{\partial }{\partial \theta} (-S(\phi ; \theta)) - \Big\langle  \frac{\partial }{\partial \theta} (-S(\phi ; \theta)) \Big\rangle_{p(\phi ; \theta)},
\end{align*}

where the term outside the expectation values is calculated on the training examples.

\subsection{$\phi^{4}$ neural network}

The case of the quantum field-theoretic neural network is more complicated due to the joint probability distribution of the visible units $\phi$ and the hidden units $h$:
\begin{equation}
p(\phi, h ; \theta) = \frac{\exp[-S(\phi,h ; \theta)]}{\int_{\bm{\phi},\bm{h}} \exp[-S(\bm{\phi},\bm{h} ; \theta)] d\bm{\phi} d\bm{h}}.
\end{equation}

From the joint probability distribution we can define marginal probability distributions via:
\begin{align*}
p(\phi ; \theta) & = \int_{\bm{h}} p(\phi, \bm{h} ; \theta) d\bm{h} = \frac{ \int_{\bm{h}} \exp[-S(\phi,\bm{h} ; \theta)] d\bm{h}}{\int_{\bm{\phi},\bm{h}} \exp[-S(\bm{\phi},\bm{h} ; \theta)] d\bm{\phi} d\bm{h}} ,
\end{align*}
\begin{align*}
 p(h ; \theta) = \int_{\bm{\phi}} p(\bm{\phi}, h ; \theta) d\bm{\phi} = \frac{ \int_{\bm{\phi}} \exp[-S(\bm{\phi},h ; \theta)] d\bm{\phi}}{\int_{\bm{\phi},\bm{h}} \exp[-S(\bm{\phi},\bm{h} ; \theta)] d\bm{\phi} d\bm{h}},
\end{align*}
as well as conditional probability distributions through:

\begin{align}
p(\phi | h ; \theta) & = \frac{p(\phi, h ; \theta)}{p(h; \theta)} = \frac{  \exp[-S(\phi,h ; \theta)] dh}{\int_{\bm{\phi}} \exp[-S(\bm{\phi},h ; \theta)] d\bm{\phi} } \\ &= \frac{\exp[\sum_{ i,j} w_{ij} \phi_{i}h_{j}  - \sum_{i} r_{i} \phi_{i} - \sum_{i} a_{i} \phi_{i}^{2} - \sum_{i} b_{i} \phi_{i}^{4}  - \cancel{\sum_{j} s_{j} h_{j}} - \cancel{\sum_{j} m_{j} h_{j}^{2}} - \cancel{\sum_{j} n_{j} h_{j}^{4}}]}{\int_{\bm{\phi}} \exp[\sum_{ i,j} w_{ij} \bm{\phi}_{i}h_{j}  - \sum_{i} r_{i} \bm{\phi}_{i} - \sum_{i} a_{i} \bm{\phi}_{i}^{2} - \sum_{i} b_{i} \bm{\phi}_{i}^{4}  - \cancel{\sum_{j} s_{j} h_{j}} - \cancel{\sum_{j} m_{j} h_{j}^{2}} - \cancel{\sum_{j} n_{j} h_{j}^{4}}]d\bm{\phi}} \\ &= \frac{\prod_{i}\exp[\phi_{i}\sum_{j} w_{ij}h_{j}  - r_{i} \phi_{i} -  a_{i} \phi_{i}^{2} - b_{i} \phi_{i}^{4} ]}{\int_{\bm{\phi}} \prod_{i} \exp[\bm{\phi}_{i}\sum_{j} w_{ij} h_{j}  -  r_{i} \bm{\phi}_{i} -  a_{i} \bm{\phi}_{i}^{2} -  b_{i} \bm{\phi}_{i}^{4}]d\bm{\phi}}  \\ &= \prod_{i}\frac{\exp[\phi_{i}\sum_{j} w_{ij}h_{j}  - r_{i} \phi_{i} -  a_{i} \phi_{i}^{2} - b_{i} \phi_{i}^{4} ]}{\int_{\bm{\phi_{i}}}  \exp[\bm{\phi}_{i}\sum_{j} w_{ij} h_{j}  -  r_{i} \bm{\phi}_{i} -  a_{i} \bm{\phi}_{i}^{2} -  b_{i} \bm{\phi}_{i}^{4}] d\bm{\phi}_{i}} \\ &= \prod_{i}p(\phi_{i} | h),
\end{align}

Similarly:
\begin{equation}
p(h | \phi ; \theta) = \frac{p(\phi, h ; \theta)}{p(\phi ; \theta)}  = \frac{ \exp[-S(\phi,h ; \theta)] }{\int_{\bm{h}} \exp[-S(\phi,\bm{h} ; \theta)] d\bm{h}} = \prod_{j} p(h_{j} | \phi).
\end{equation}

The gradient of the log-likelihood for the case of the quantum field-theoretic neural network is:
\begin{align}
\frac{\partial \ln p(\phi ; \theta) }{ \partial \theta} &= \frac{\partial }{\partial \theta} \Bigg[ \ln \frac{\int_{\bm{h}} \exp [-S(\phi,\bm{h} ; \theta)]d\bm{h}}{\int_{\bm{\phi},\bm{h}} \exp [-S(\bm{\phi},\bm{h} ; \theta)] d\bm{\phi} d\bm{h}}  \Bigg] \\ &= \frac{\partial }{\partial \theta} \Bigg[ \ln \int_{\bm{h}} \exp [-S(\phi,\bm{h} ; \theta)]d\bm{h} - \ln\int_{\bm{\phi},\bm{h}} \exp [-S(\bm{\phi},\bm{h} ; \theta)] d\bm{\phi} d\bm{h}  \Bigg] \\ &= \frac{\int_{\bm{h}}\frac{\partial }{\partial \theta} (-S(\phi,\bm{h} ; \theta))  \exp [-S(\phi,\bm{h} ; \theta)]d\bm{h}  }{\int_{\bm{h}} \exp [-S(\phi,\bm{h} ; \theta)]d\bm{h} } - \frac{\int_{\bm{\phi}, \bm{h}}\frac{\partial }{\partial \theta} (-S(\bm{\phi},\bm{h} ; \theta))  \exp [-S(\bm{\phi},\bm{h} ; \theta)] d\bm{\phi} d\bm{h}  }{\int_{\bm{\phi}, \bm{h}} \exp [-S(\bm{\phi},\bm{h} ; \theta)]d\bm{\phi} d\bm{h} } \\ &= \int_{\bm{h}} p(\bm{h}| \phi ; \theta) \frac{\partial }{\partial \theta} (-S(\phi,\bm{h} ; \theta)) d\bm{h} - \int_{\bm{\phi},\bm{h}} p(\bm{\phi}, \bm{h} ; \theta) \frac{\partial }{\partial \theta} (-S(\bm{\phi},\bm{h} ; \theta))  d\bm{\phi} d\bm{h} \\ &= \Big\langle \frac{\partial }{\partial \theta} (-S(\phi,h ; \theta)) \Big\rangle_{p(h | \phi ; \theta)} - \Big\langle \frac{\partial }{\partial \theta} (-S(\phi,h ; \theta)) \Big\rangle_{p(\phi, h ; \theta)}. 
\end{align}

We approximate the last expression in the above equation for each parameter $\theta$ using contrastive divergence. Specifically, the visible units $\phi$ are set equal to a specific training example $\phi^{(x)}$ and then based on the conditional distribution $p(h | \phi^{(x)})$ a set of hidden units $h^{(x)}$ is sampled. The hidden units $h^{(x)}$ are then utilized to sample a new set of visible units $\phi^{(x+1)}$ and the approach is repeated for $k$ steps:
\begin{equation}
CD_{k}=  \Big\langle\frac{\partial }{\partial \theta} (-S(\phi^{(0)},h ; \theta))\Big\rangle_{p(h | \phi^{(0)} ; \theta)} - \Big\langle  \frac{\partial }{\partial \theta} (-S(\phi^{(k)},h ; \theta)) \Big\rangle_{p( h | \phi^{(k)} ; \theta)},
\end{equation}
where for the considered cases we use $k=1$. 
\end{widetext}

\section{Simulation details and hyperparameters \label{app:sim}}

The $\phi^{4}$ scalar field theory is a system with continuous degrees of freedom $-\infty < \phi_{i} < +\infty$. To sample the system we implement Markov chain Monte Carlo sampling with the Metropolis algorithm, where we consider one step as equivalent to updating a number of lattice sites equal to the volume of the system.  The question of how to properly choose a new state additionally arises. When the training data have values which lie at a specific interval, the aim of the machine learning algorithm is to learn a probability distribution which reproduces them. The new state can then be chosen by sampling uniformly between the minimum and maximum value, therefore guaranteeing that every state is reachable under an arbitrary large number of Monte Carlo steps. For the case of the hidden units in the $\phi^{4}$ neural network we impose the same restriction, even though the hidden units could, in principle, remain unconstrained, i.e. $-\infty < h < +\infty$. We also emphasize that during the gradient process of the Markov random field we retain one Markov chain to conduct the necessary calculations.

The learning rate that produced Figs.~\ref{fig:heat} and~\ref{fig:kl} is $10^{-3}$ and $10^{-2}$, respectively. The sample size is chosen equal to $50$ before updating the variational parameters $\theta$. The image in Fig.~\ref{fig:bird} has size $32*32$ and its continuous values lie between $[-1,1]$. The Markov random field was trained with learning rate $0.1$ and $4 \times 10^{4}$ epochs. The parameters that produced Fig.~\ref{fig:2} are a learning rate of $0.1$, $400$ epochs and a batch size of $4$. For the results depicted in Fig.~\ref{fig:faces} the $\phi^{4}$ neural network has $4096$ visible units, $32$ hidden units, learning rate $0.1$,  batch size of $5$ and was trained for $10^{4}$ epochs on the first $40$ examples of the Olivetti faces data set.

\section{Histogram reweighting \label{app:reweig}}

We consider the numerical estimator for an arbitrary observable $\langle O \rangle$ in the full complex action $\mathcal{A}$ which we aim to sample:
\begin{equation}\label{eq:apprew}
\langle O \rangle= \frac{\sum_{l=1}^{N} O_{{l}} \tilde{p_{{l}}}^{-1} \exp[-g_{j}'\mathcal{A}_{{l}}^{(j)}- \sum_{k=1,k \neq j}^{5}g_{k}\mathcal{A}^{(k)}_{{l}}]}{\sum_{l=1}^{N} \tilde{p_{{l}}}^{-1}  \exp[-g_{j}'\mathcal{A}_{{l}}^{(j)}- \sum_{k=1,k \neq j}^{5}g_{k}\mathcal{A}_{{l}}^{(k)}]},
\end{equation}
where $N$ is the subset of Monte Carlo samples and $\tilde{p}$ are the probabilities used to sample from the equilibrium distribution. We have expressed the numerical estimator in a form that simultaneously allows extrapolation along the trajectory of a coupling constant $g_{j}'$. We will now substitute the probabilities $\tilde{p}$ for the probabilities of the inhomogeneous $\phi^{4}$ Markov random field:
\begin{equation}
\tilde{p_{{l}}}= \frac{\exp[-S_{{l}}]}{\int_{\bm{\phi}} \exp[-S_{\bm{\phi}}] d\bm{\phi}},
\end{equation}
where the sum is over all possible states $\bm{\phi}$ of the system and we arrive at the reweighting equation:
\begin{equation}
\langle O \rangle= \frac{\sum_{l=1}^{N} O_{l} \exp[S_{l}-g_{j}'\mathcal{A}_{l}^{(j)}- \sum_{k=1,k \neq j}^{5}g_{k}\mathcal{A}^{(k)}_{l}]}{\sum_{l=1}^{N}  \exp[S_{l}-g_{j}'\mathcal{A}_{l}^{(j)}- \sum_{k=1,k \neq j}^{5}g_{k}\mathcal{A}_{l}^{(k)}]}.
\end{equation}

Given a subset of samples drawn from the equilibrium distribution of the $\phi^{4}$ Markov random field, which is described by the action $S$, one can extrapolate observables to the full distribution of the action $\mathcal{A}$ which includes longer range interactions and complex-valued terms along the trajectory of a coupling constant $g_{j}'$.

To compare the reweighting extrapolations from the $\phi^{4}$ Markov random field to the full action, we additionally implement reweighting from the simulated action $A_{\{4\}}$. In this form of reweighting we consider again Eq.~\ref{eq:apprew} and we substitute $\tilde{p}$ for:
\begin{equation}
\tilde{p_{l}}= \frac{\exp[- \sum_{k=1}^{4}g_{k}\mathcal{A}_{l}^{(k)}]}{\int_{\bm{\phi}} \exp[- \sum_{k=1}^{4}g_{k}\mathcal{A}_{\bm{\phi}}^{(k)}] d\bm{\phi}},
\end{equation}
where we consider for this specific case that $g_{j}'=g_{j}$, arriving at the equation:

\begin{equation}
\langle O \rangle= \frac{\sum_{l=1}^{N} O_{l} \exp[-\Im{A_{l}}]}{\sum_{l=1}^{N}  \exp[-\Im{A_{l}}]}.
\end{equation}
One observable of interest is the magnetization which is defined as:
\begin{equation}
m= \frac{1}{V} \Bigg| \sum_{i} \phi_{i} \Bigg|,
\end{equation}
where $V=L*L$ is the volume of the system.

\section{Binning analysis \label{app:err}}

Statistical errors are calculated with the binning method on the obtained Monte Carlo data sets. Each data set with $10^{4}$ minimally correlated configurations is split into $n=10$ data sets where calculations of observables $O$ are conducted. The standard deviation for an observable $O$ is then obtained through:

\begin{equation}
\sigma_{O}= \sqrt{\frac{1}{n-1} (\overline{O^{2}}-\overline{O}^{2})}.
\end{equation}

\bibliography{ms}

\providecommand{\noopsort}[1]{}\providecommand{\singleletter}[1]{#1}%
\begin{thebibliography}{65}%
\makeatletter
\providecommand \@ifxundefined [1]{%
 \@ifx{#1\undefined}
}%
\providecommand \@ifnum [1]{%
 \ifnum #1\expandafter \@firstoftwo
 \else \expandafter \@secondoftwo
 \fi
}%
\providecommand \@ifx [1]{%
 \ifx #1\expandafter \@firstoftwo
 \else \expandafter \@secondoftwo
 \fi
}%
\providecommand \natexlab [1]{#1}%
\providecommand \enquote  [1]{``#1''}%
\providecommand \bibnamefont  [1]{#1}%
\providecommand \bibfnamefont [1]{#1}%
\providecommand \citenamefont [1]{#1}%
\providecommand \href@noop [0]{\@secondoftwo}%
\providecommand \href [0]{\begingroup \@sanitize@url \@href}%
\providecommand \@href[1]{\@@startlink{#1}\@@href}%
\providecommand \@@href[1]{\endgroup#1\@@endlink}%
\providecommand \@sanitize@url [0]{\catcode `\\12\catcode `\$12\catcode
  `\&12\catcode `\#12\catcode `\^12\catcode `\_12\catcode `\%12\relax}%
\providecommand \@@startlink[1]{}%
\providecommand \@@endlink[0]{}%
\providecommand \url  [0]{\begingroup\@sanitize@url \@url }%
\providecommand \@url [1]{\endgroup\@href {#1}{\urlprefix }}%
\providecommand \urlprefix  [0]{URL }%
\providecommand \Eprint [0]{\href }%
\providecommand \doibase [0]{https://doi.org/}%
\providecommand \selectlanguage [0]{\@gobble}%
\providecommand \bibinfo  [0]{\@secondoftwo}%
\providecommand \bibfield  [0]{\@secondoftwo}%
\providecommand \translation [1]{[#1]}%
\providecommand \BibitemOpen [0]{}%
\providecommand \bibitemStop [0]{}%
\providecommand \bibitemNoStop [0]{.\EOS\space}%
\providecommand \EOS [0]{\spacefactor3000\relax}%
\providecommand \BibitemShut  [1]{\csname bibitem#1\endcsname}%
\let\auto@bib@innerbib\@empty
\bibitem [{\citenamefont {Zinn-Justin}(2002)}]{jean}%
  \BibitemOpen
  \bibfield  {author} {\bibinfo {author} {\bibfnamefont {J.}~\bibnamefont
  {Zinn-Justin}},\ }\href@noop {} {\emph {\bibinfo {title} {Quantum Field
  Theory and Critical Phenomena}}}\ (\bibinfo  {publisher} {Oxford University
  Press},\ \bibinfo {address} {Oxford},\ \bibinfo {year} {2002})\BibitemShut
  {NoStop}%
\bibitem [{\citenamefont {Glimm}\ and\ \citenamefont {Jaffe}(1987)}]{glimmcft}%
  \BibitemOpen
  \bibfield  {author} {\bibinfo {author} {\bibfnamefont {J.}~\bibnamefont
  {Glimm}}\ and\ \bibinfo {author} {\bibfnamefont {A.}~\bibnamefont {Jaffe}},\
  }\href@noop {} {\emph {\bibinfo {title} {Quantum Physics: A Functional
  Integral Point of View}}}\ (\bibinfo  {publisher} {Springer, New York, NY},\
  \bibinfo {year} {1987})\BibitemShut {NoStop}%
\bibitem [{\citenamefont {Velo}\ and\ \citenamefont
  {Wightman}(1973)}]{velo1973constructive}%
  \BibitemOpen
  \bibfield  {author} {\bibinfo {author} {\bibfnamefont {G.}~\bibnamefont
  {Velo}}\ and\ \bibinfo {author} {\bibfnamefont {A.}~\bibnamefont
  {Wightman}},\ }\href@noop {} {\emph {\bibinfo {title} {Constructive Quantum
  Field Theory: The 1973 “Ettore Majorana” International School of
  Mathematical Physics}}},\ Lecture Notes in Physics\ (\bibinfo  {publisher}
  {Springer Berlin Heidelberg},\ \bibinfo {year} {1973})\BibitemShut {NoStop}%
\bibitem [{\citenamefont {Seiler}(1982)}]{seiler}%
  \BibitemOpen
  \bibfield  {author} {\bibinfo {author} {\bibfnamefont {E.}~\bibnamefont
  {Seiler}},\ }\href@noop {} {\emph {\bibinfo {title} {Gauge Theories as a
  Problem of Constructive Quantum Field Theory and Statistical Mechanics}}}\
  (\bibinfo  {publisher} {Springer, Berlin, Heidelberg},\ \bibinfo {year}
  {1982})\BibitemShut {NoStop}%
\bibitem [{\citenamefont {Nelson}(1973{\natexlab{a}})}]{Nelson1973}%
  \BibitemOpen
  \bibfield  {author} {\bibinfo {author} {\bibfnamefont {E.}~\bibnamefont
  {Nelson}},\ }\bibinfo {title} {Probability theory and euclidean field
  theory},\ in\ \href {https://doi.org/10.1007/BFb0113084} {\emph {\bibinfo
  {booktitle} {Constructive Quantum Field Theory}}},\ \bibinfo {editor} {edited
  by\ \bibinfo {editor} {\bibfnamefont {G.}~\bibnamefont {Velo}}\ and\ \bibinfo
  {editor} {\bibfnamefont {A.}~\bibnamefont {Wightman}}}\ (\bibinfo
  {publisher} {Springer Berlin Heidelberg},\ \bibinfo {address} {Berlin,
  Heidelberg},\ \bibinfo {year} {1973})\ pp.\ \bibinfo {pages}
  {94--124}\BibitemShut {NoStop}%
\bibitem [{\citenamefont {Nelson}(1973{\natexlab{b}})}]{NELSON197397}%
  \BibitemOpen
  \bibfield  {author} {\bibinfo {author} {\bibfnamefont {E.}~\bibnamefont
  {Nelson}},\ }\bibfield  {title} {\bibinfo {title} {Construction of quantum
  fields from markoff fields},\ }\href
  {https://doi.org/https://doi.org/10.1016/0022-1236(73)90091-8} {\bibfield
  {journal} {\bibinfo  {journal} {Journal of Functional Analysis}\ }\textbf
  {\bibinfo {volume} {12}},\ \bibinfo {pages} {97 } (\bibinfo {year}
  {1973}{\natexlab{b}})}\BibitemShut {NoStop}%
\bibitem [{\citenamefont {Goodfellow}\ \emph {et~al.}(2016)\citenamefont
  {Goodfellow}, \citenamefont {Bengio},\ and\ \citenamefont
  {Courville}}]{GoodBengCour16}%
  \BibitemOpen
  \bibfield  {author} {\bibinfo {author} {\bibfnamefont {I.~J.}\ \bibnamefont
  {Goodfellow}}, \bibinfo {author} {\bibfnamefont {Y.}~\bibnamefont {Bengio}},\
  and\ \bibinfo {author} {\bibfnamefont {A.}~\bibnamefont {Courville}},\
  }\href@noop {} {\emph {\bibinfo {title} {Deep Learning}}}\ (\bibinfo
  {publisher} {MIT Press},\ \bibinfo {address} {Cambridge, MA, USA},\ \bibinfo
  {year} {2016})\ \bibinfo {note}
  {\url{http://www.deeplearningbook.org}}\BibitemShut {NoStop}%
\bibitem [{\citenamefont {Carleo}\ \emph {et~al.}(2019)\citenamefont {Carleo},
  \citenamefont {Cirac}, \citenamefont {Cranmer}, \citenamefont {Daudet},
  \citenamefont {Schuld}, \citenamefont {Tishby}, \citenamefont
  {Vogt-Maranto},\ and\ \citenamefont {Zdeborov\'a}}]{Carleo_2019}%
  \BibitemOpen
  \bibfield  {author} {\bibinfo {author} {\bibfnamefont {G.}~\bibnamefont
  {Carleo}}, \bibinfo {author} {\bibfnamefont {I.}~\bibnamefont {Cirac}},
  \bibinfo {author} {\bibfnamefont {K.}~\bibnamefont {Cranmer}}, \bibinfo
  {author} {\bibfnamefont {L.}~\bibnamefont {Daudet}}, \bibinfo {author}
  {\bibfnamefont {M.}~\bibnamefont {Schuld}}, \bibinfo {author} {\bibfnamefont
  {N.}~\bibnamefont {Tishby}}, \bibinfo {author} {\bibfnamefont
  {L.}~\bibnamefont {Vogt-Maranto}},\ and\ \bibinfo {author} {\bibfnamefont
  {L.}~\bibnamefont {Zdeborov\'a}},\ }\bibfield  {title} {\bibinfo {title}
  {Machine learning and the physical sciences},\ }\bibfield  {journal}
  {\bibinfo  {journal} {Reviews of Modern Physics}\ }\textbf {\bibinfo {volume}
  {91}},\ \href {https://doi.org/10.1103/revmodphys.91.045002}
  {10.1103/revmodphys.91.045002} (\bibinfo {year} {2019})\BibitemShut {NoStop}%
\bibitem [{\citenamefont {Kanwar}\ \emph {et~al.}(2020)\citenamefont {Kanwar},
  \citenamefont {Albergo}, \citenamefont {Boyda}, \citenamefont {Cranmer},
  \citenamefont {Hackett}, \citenamefont {Racani\`ere}, \citenamefont
  {Rezende},\ and\ \citenamefont {Shanahan}}]{PhysRevLett.125.121601}%
  \BibitemOpen
  \bibfield  {author} {\bibinfo {author} {\bibfnamefont {G.}~\bibnamefont
  {Kanwar}}, \bibinfo {author} {\bibfnamefont {M.~S.}\ \bibnamefont {Albergo}},
  \bibinfo {author} {\bibfnamefont {D.}~\bibnamefont {Boyda}}, \bibinfo
  {author} {\bibfnamefont {K.}~\bibnamefont {Cranmer}}, \bibinfo {author}
  {\bibfnamefont {D.~C.}\ \bibnamefont {Hackett}}, \bibinfo {author}
  {\bibfnamefont {S.}~\bibnamefont {Racani\`ere}}, \bibinfo {author}
  {\bibfnamefont {D.~J.}\ \bibnamefont {Rezende}},\ and\ \bibinfo {author}
  {\bibfnamefont {P.~E.}\ \bibnamefont {Shanahan}},\ }\bibfield  {title}
  {\bibinfo {title} {Equivariant flow-based sampling for lattice gauge
  theory},\ }\href {https://doi.org/10.1103/PhysRevLett.125.121601} {\bibfield
  {journal} {\bibinfo  {journal} {Phys. Rev. Lett.}\ }\textbf {\bibinfo
  {volume} {125}},\ \bibinfo {pages} {121601} (\bibinfo {year}
  {2020})}\BibitemShut {NoStop}%
\bibitem [{\citenamefont {Shanahan}\ \emph {et~al.}(2018)\citenamefont
  {Shanahan}, \citenamefont {Trewartha},\ and\ \citenamefont
  {Detmold}}]{PhysRevD.97.094506}%
  \BibitemOpen
  \bibfield  {author} {\bibinfo {author} {\bibfnamefont {P.~E.}\ \bibnamefont
  {Shanahan}}, \bibinfo {author} {\bibfnamefont {D.}~\bibnamefont
  {Trewartha}},\ and\ \bibinfo {author} {\bibfnamefont {W.}~\bibnamefont
  {Detmold}},\ }\bibfield  {title} {\bibinfo {title} {Machine learning action
  parameters in lattice quantum chromodynamics},\ }\href
  {https://doi.org/10.1103/PhysRevD.97.094506} {\bibfield  {journal} {\bibinfo
  {journal} {Phys. Rev. D}\ }\textbf {\bibinfo {volume} {97}},\ \bibinfo
  {pages} {094506} (\bibinfo {year} {2018})}\BibitemShut {NoStop}%
\bibitem [{\citenamefont {Zhou}\ \emph {et~al.}(2019)\citenamefont {Zhou},
  \citenamefont {Endr\ifmmode~\mbox{\H{o}}\else \H{o}\fi{}di}, \citenamefont
  {Pang},\ and\ \citenamefont {St\"ocker}}]{PhysRevD.100.011501}%
  \BibitemOpen
  \bibfield  {author} {\bibinfo {author} {\bibfnamefont {K.}~\bibnamefont
  {Zhou}}, \bibinfo {author} {\bibfnamefont {G.}~\bibnamefont
  {Endr\ifmmode~\mbox{\H{o}}\else \H{o}\fi{}di}}, \bibinfo {author}
  {\bibfnamefont {L.-G.}\ \bibnamefont {Pang}},\ and\ \bibinfo {author}
  {\bibfnamefont {H.}~\bibnamefont {St\"ocker}},\ }\bibfield  {title} {\bibinfo
  {title} {Regressive and generative neural networks for scalar field theory},\
  }\href {https://doi.org/10.1103/PhysRevD.100.011501} {\bibfield  {journal}
  {\bibinfo  {journal} {Phys. Rev. D}\ }\textbf {\bibinfo {volume} {100}},\
  \bibinfo {pages} {011501} (\bibinfo {year} {2019})}\BibitemShut {NoStop}%
\bibitem [{\citenamefont {Bachtis}\ \emph
  {et~al.}(2020{\natexlab{a}})\citenamefont {Bachtis}, \citenamefont {Aarts},\
  and\ \citenamefont {Lucini}}]{PhysRevE.102.053306}%
  \BibitemOpen
  \bibfield  {author} {\bibinfo {author} {\bibfnamefont {D.}~\bibnamefont
  {Bachtis}}, \bibinfo {author} {\bibfnamefont {G.}~\bibnamefont {Aarts}},\
  and\ \bibinfo {author} {\bibfnamefont {B.}~\bibnamefont {Lucini}},\
  }\bibfield  {title} {\bibinfo {title} {Mapping distinct phase transitions to
  a neural network},\ }\href {https://doi.org/10.1103/PhysRevE.102.053306}
  {\bibfield  {journal} {\bibinfo  {journal} {Phys. Rev. E}\ }\textbf {\bibinfo
  {volume} {102}},\ \bibinfo {pages} {053306} (\bibinfo {year}
  {2020}{\natexlab{a}})}\BibitemShut {NoStop}%
\bibitem [{\citenamefont {Chernodub}\ \emph {et~al.}(2020)\citenamefont
  {Chernodub}, \citenamefont {Erbin}, \citenamefont {Goy},\ and\ \citenamefont
  {Molochkov}}]{PhysRevD.102.054501}%
  \BibitemOpen
  \bibfield  {author} {\bibinfo {author} {\bibfnamefont {M.~N.}\ \bibnamefont
  {Chernodub}}, \bibinfo {author} {\bibfnamefont {H.}~\bibnamefont {Erbin}},
  \bibinfo {author} {\bibfnamefont {V.~A.}\ \bibnamefont {Goy}},\ and\ \bibinfo
  {author} {\bibfnamefont {A.~V.}\ \bibnamefont {Molochkov}},\ }\bibfield
  {title} {\bibinfo {title} {Topological defects and confinement with machine
  learning: The case of monopoles in compact electrodynamics},\ }\href
  {https://doi.org/10.1103/PhysRevD.102.054501} {\bibfield  {journal} {\bibinfo
   {journal} {Phys. Rev. D}\ }\textbf {\bibinfo {volume} {102}},\ \bibinfo
  {pages} {054501} (\bibinfo {year} {2020})}\BibitemShut {NoStop}%
\bibitem [{\citenamefont {Bl\"ucher}\ \emph {et~al.}(2020)\citenamefont
  {Bl\"ucher}, \citenamefont {Kades}, \citenamefont {Pawlowski}, \citenamefont
  {Strodthoff},\ and\ \citenamefont {Urban}}]{PhysRevD.101.094507}%
  \BibitemOpen
  \bibfield  {author} {\bibinfo {author} {\bibfnamefont {S.}~\bibnamefont
  {Bl\"ucher}}, \bibinfo {author} {\bibfnamefont {L.}~\bibnamefont {Kades}},
  \bibinfo {author} {\bibfnamefont {J.~M.}\ \bibnamefont {Pawlowski}}, \bibinfo
  {author} {\bibfnamefont {N.}~\bibnamefont {Strodthoff}},\ and\ \bibinfo
  {author} {\bibfnamefont {J.~M.}\ \bibnamefont {Urban}},\ }\bibfield  {title}
  {\bibinfo {title} {Towards novel insights in lattice field theory with
  explainable machine learning},\ }\href
  {https://doi.org/10.1103/PhysRevD.101.094507} {\bibfield  {journal} {\bibinfo
   {journal} {Phys. Rev. D}\ }\textbf {\bibinfo {volume} {101}},\ \bibinfo
  {pages} {094507} (\bibinfo {year} {2020})}\BibitemShut {NoStop}%
\bibitem [{\citenamefont {Favoni}\ \emph {et~al.}(2020)\citenamefont {Favoni},
  \citenamefont {Ipp}, \citenamefont {Müller},\ and\ \citenamefont
  {Schuh}}]{favoni2020lattice}%
  \BibitemOpen
  \bibfield  {author} {\bibinfo {author} {\bibfnamefont {M.}~\bibnamefont
  {Favoni}}, \bibinfo {author} {\bibfnamefont {A.}~\bibnamefont {Ipp}},
  \bibinfo {author} {\bibfnamefont {D.~I.}\ \bibnamefont {Müller}},\ and\
  \bibinfo {author} {\bibfnamefont {D.}~\bibnamefont {Schuh}},\ }\href@noop {}
  {\bibinfo {title} {Lattice gauge equivariant convolutional neural networks}}
  (\bibinfo {year} {2020}),\ \Eprint {https://arxiv.org/abs/2012.12901}
  {arXiv:2012.12901 [hep-lat]} \BibitemShut {NoStop}%
\bibitem [{\citenamefont {Nicoli}\ \emph {et~al.}(2021)\citenamefont {Nicoli},
  \citenamefont {Anders}, \citenamefont {Funcke}, \citenamefont {Hartung},
  \citenamefont {Jansen}, \citenamefont {Kessel}, \citenamefont {Nakajima},\
  and\ \citenamefont {Stornati}}]{nicoli2021estimation}%
  \BibitemOpen
  \bibfield  {author} {\bibinfo {author} {\bibfnamefont {K.~A.}\ \bibnamefont
  {Nicoli}}, \bibinfo {author} {\bibfnamefont {C.~J.}\ \bibnamefont {Anders}},
  \bibinfo {author} {\bibfnamefont {L.}~\bibnamefont {Funcke}}, \bibinfo
  {author} {\bibfnamefont {T.}~\bibnamefont {Hartung}}, \bibinfo {author}
  {\bibfnamefont {K.}~\bibnamefont {Jansen}}, \bibinfo {author} {\bibfnamefont
  {P.}~\bibnamefont {Kessel}}, \bibinfo {author} {\bibfnamefont
  {S.}~\bibnamefont {Nakajima}},\ and\ \bibinfo {author} {\bibfnamefont
  {P.}~\bibnamefont {Stornati}},\ }\href@noop {} {\bibinfo {title} {Estimation
  of thermodynamic observables in lattice field theories with deep generative
  models}} (\bibinfo {year} {2021}),\ \Eprint
  {https://arxiv.org/abs/2007.07115} {arXiv:2007.07115 [hep-lat]} \BibitemShut
  {NoStop}%
\bibitem [{\citenamefont {Hu}\ \emph {et~al.}(2020)\citenamefont {Hu},
  \citenamefont {Li}, \citenamefont {Wang},\ and\ \citenamefont
  {You}}]{PhysRevResearch.2.023369}%
  \BibitemOpen
  \bibfield  {author} {\bibinfo {author} {\bibfnamefont {H.-Y.}\ \bibnamefont
  {Hu}}, \bibinfo {author} {\bibfnamefont {S.-H.}\ \bibnamefont {Li}}, \bibinfo
  {author} {\bibfnamefont {L.}~\bibnamefont {Wang}},\ and\ \bibinfo {author}
  {\bibfnamefont {Y.-Z.}\ \bibnamefont {You}},\ }\bibfield  {title} {\bibinfo
  {title} {Machine learning holographic mapping by neural network
  renormalization group},\ }\href
  {https://doi.org/10.1103/PhysRevResearch.2.023369} {\bibfield  {journal}
  {\bibinfo  {journal} {Phys. Rev. Research}\ }\textbf {\bibinfo {volume}
  {2}},\ \bibinfo {pages} {023369} (\bibinfo {year} {2020})}\BibitemShut
  {NoStop}%
\bibitem [{\citenamefont {van Nieuwenburg}\ \emph {et~al.}(2017)\citenamefont
  {van Nieuwenburg}, \citenamefont {Liu},\ and\ \citenamefont
  {Huber}}]{vanNieuwenburg2017}%
  \BibitemOpen
  \bibfield  {author} {\bibinfo {author} {\bibfnamefont {E.~L.}\ \bibnamefont
  {van Nieuwenburg}}, \bibinfo {author} {\bibfnamefont {Y.-H.}\ \bibnamefont
  {Liu}},\ and\ \bibinfo {author} {\bibfnamefont {S.}~\bibnamefont {Huber}},\
  }\bibfield  {title} {\bibinfo {title} {Learning phase transitions by
  confusion},\ }\href {https://doi.org/10.1038/nphys4037} {\bibfield  {journal}
  {\bibinfo  {journal} {Nature Physics}\ }\textbf {\bibinfo {volume} {13}},\
  \bibinfo {pages} {435} (\bibinfo {year} {2017})}\BibitemShut {NoStop}%
\bibitem [{\citenamefont {Carrasquilla}\ and\ \citenamefont
  {Melko}(2017)}]{Carrasquilla2017}%
  \BibitemOpen
  \bibfield  {author} {\bibinfo {author} {\bibfnamefont {J.}~\bibnamefont
  {Carrasquilla}}\ and\ \bibinfo {author} {\bibfnamefont {R.~G.}\ \bibnamefont
  {Melko}},\ }\bibfield  {title} {\bibinfo {title} {Machine learning phases of
  matter},\ }\href {https://doi.org/10.1038/nphys4035} {\bibfield  {journal}
  {\bibinfo  {journal} {Nature Physics}\ }\textbf {\bibinfo {volume} {13}},\
  \bibinfo {pages} {431} (\bibinfo {year} {2017})}\BibitemShut {NoStop}%
\bibitem [{\citenamefont {Giannetti}\ \emph {et~al.}(2019)\citenamefont
  {Giannetti}, \citenamefont {Lucini},\ and\ \citenamefont
  {Vadacchino}}]{GIANNETTI2019114639}%
  \BibitemOpen
  \bibfield  {author} {\bibinfo {author} {\bibfnamefont {C.}~\bibnamefont
  {Giannetti}}, \bibinfo {author} {\bibfnamefont {B.}~\bibnamefont {Lucini}},\
  and\ \bibinfo {author} {\bibfnamefont {D.}~\bibnamefont {Vadacchino}},\
  }\bibfield  {title} {\bibinfo {title} {Machine learning as a universal tool
  for quantitative investigations of phase transitions},\ }\href
  {https://doi.org/https://doi.org/10.1016/j.nuclphysb.2019.114639} {\bibfield
  {journal} {\bibinfo  {journal} {Nuclear Physics B}\ }\textbf {\bibinfo
  {volume} {944}},\ \bibinfo {pages} {114639} (\bibinfo {year}
  {2019})}\BibitemShut {NoStop}%
\bibitem [{\citenamefont {Bachtis}\ \emph {et~al.}(2021)\citenamefont
  {Bachtis}, \citenamefont {Aarts},\ and\ \citenamefont
  {Lucini}}]{bachtis2020adding}%
  \BibitemOpen
  \bibfield  {author} {\bibinfo {author} {\bibfnamefont {D.}~\bibnamefont
  {Bachtis}}, \bibinfo {author} {\bibfnamefont {G.}~\bibnamefont {Aarts}},\
  and\ \bibinfo {author} {\bibfnamefont {B.}~\bibnamefont {Lucini}},\
  }\bibfield  {title} {\bibinfo {title} {Adding machine learning within
  hamiltonians: Renormalization group transformations, symmetry breaking and
  restoration},\ }\href {https://doi.org/10.1103/PhysRevResearch.3.013134}
  {\bibfield  {journal} {\bibinfo  {journal} {Phys. Rev. Research}\ }\textbf
  {\bibinfo {volume} {3}},\ \bibinfo {pages} {013134} (\bibinfo {year}
  {2021})}\BibitemShut {NoStop}%
\bibitem [{\citenamefont {Bachtis}\ \emph
  {et~al.}(2020{\natexlab{b}})\citenamefont {Bachtis}, \citenamefont {Aarts},\
  and\ \citenamefont {Lucini}}]{bachtis2020extending}%
  \BibitemOpen
  \bibfield  {author} {\bibinfo {author} {\bibfnamefont {D.}~\bibnamefont
  {Bachtis}}, \bibinfo {author} {\bibfnamefont {G.}~\bibnamefont {Aarts}},\
  and\ \bibinfo {author} {\bibfnamefont {B.}~\bibnamefont {Lucini}},\
  }\bibfield  {title} {\bibinfo {title} {Extending machine learning
  classification capabilities with histogram reweighting},\ }\href
  {https://doi.org/10.1103/PhysRevE.102.033303} {\bibfield  {journal} {\bibinfo
   {journal} {Phys. Rev. E}\ }\textbf {\bibinfo {volume} {102}},\ \bibinfo
  {pages} {033303} (\bibinfo {year} {2020}{\natexlab{b}})}\BibitemShut
  {NoStop}%
\bibitem [{\citenamefont {Wang}(2016)}]{PhysRevB.94.195105}%
  \BibitemOpen
  \bibfield  {author} {\bibinfo {author} {\bibfnamefont {L.}~\bibnamefont
  {Wang}},\ }\bibfield  {title} {\bibinfo {title} {Discovering phase
  transitions with unsupervised learning},\ }\href
  {https://doi.org/10.1103/PhysRevB.94.195105} {\bibfield  {journal} {\bibinfo
  {journal} {Phys. Rev. B}\ }\textbf {\bibinfo {volume} {94}},\ \bibinfo
  {pages} {195105} (\bibinfo {year} {2016})}\BibitemShut {NoStop}%
\bibitem [{\citenamefont {Tanaka}\ and\ \citenamefont
  {Tomiya}(2017)}]{doi:10.7566/JPSJ.86.063001}%
  \BibitemOpen
  \bibfield  {author} {\bibinfo {author} {\bibfnamefont {A.}~\bibnamefont
  {Tanaka}}\ and\ \bibinfo {author} {\bibfnamefont {A.}~\bibnamefont
  {Tomiya}},\ }\bibfield  {title} {\bibinfo {title} {Detection of phase
  transition via convolutional neural networks},\ }\href
  {https://doi.org/10.7566/JPSJ.86.063001} {\bibfield  {journal} {\bibinfo
  {journal} {Journal of the Physical Society of Japan}\ }\textbf {\bibinfo
  {volume} {86}},\ \bibinfo {pages} {063001} (\bibinfo {year} {2017})},\
  \Eprint {https://arxiv.org/abs/https://doi.org/10.7566/JPSJ.86.063001}
  {https://doi.org/10.7566/JPSJ.86.063001} \BibitemShut {NoStop}%
\bibitem [{\citenamefont {Agliari}\ \emph {et~al.}(2020)\citenamefont
  {Agliari}, \citenamefont {Barra}, \citenamefont {Sollich},\ and\
  \citenamefont {Zdeborov{\'{a}}}}]{Agliari_2020}%
  \BibitemOpen
  \bibfield  {author} {\bibinfo {author} {\bibfnamefont {E.}~\bibnamefont
  {Agliari}}, \bibinfo {author} {\bibfnamefont {A.}~\bibnamefont {Barra}},
  \bibinfo {author} {\bibfnamefont {P.}~\bibnamefont {Sollich}},\ and\ \bibinfo
  {author} {\bibfnamefont {L.}~\bibnamefont {Zdeborov{\'{a}}}},\ }\bibfield
  {title} {\bibinfo {title} {Machine learning and statistical physics:
  preface},\ }\href {https://doi.org/10.1088/1751-8121/abca75} {\bibfield
  {journal} {\bibinfo  {journal} {Journal of Physics A: Mathematical and
  Theoretical}\ }\textbf {\bibinfo {volume} {53}},\ \bibinfo {pages} {500401}
  (\bibinfo {year} {2020})}\BibitemShut {NoStop}%
\bibitem [{\citenamefont {Zdeborova}\ and\ \citenamefont
  {Krzakala}(2016)}]{doi:10.1080/00018732.2016.1211393}%
  \BibitemOpen
  \bibfield  {author} {\bibinfo {author} {\bibfnamefont {L.}~\bibnamefont
  {Zdeborova}}\ and\ \bibinfo {author} {\bibfnamefont {F.}~\bibnamefont
  {Krzakala}},\ }\bibfield  {title} {\bibinfo {title} {Statistical physics of
  inference: thresholds and algorithms},\ }\href
  {https://doi.org/10.1080/00018732.2016.1211393} {\bibfield  {journal}
  {\bibinfo  {journal} {Advances in Physics}\ }\textbf {\bibinfo {volume}
  {65}},\ \bibinfo {pages} {453} (\bibinfo {year} {2016})},\ \Eprint
  {https://arxiv.org/abs/https://doi.org/10.1080/00018732.2016.1211393}
  {https://doi.org/10.1080/00018732.2016.1211393} \BibitemShut {NoStop}%
\bibitem [{\citenamefont {Goldt}\ \emph {et~al.}(2020)\citenamefont {Goldt},
  \citenamefont {Advani}, \citenamefont {Saxe}, \citenamefont {Krzakala},\ and\
  \citenamefont {Zdeborov{\'{a}}}}]{Goldt_2020}%
  \BibitemOpen
  \bibfield  {author} {\bibinfo {author} {\bibfnamefont {S.}~\bibnamefont
  {Goldt}}, \bibinfo {author} {\bibfnamefont {M.~S.}\ \bibnamefont {Advani}},
  \bibinfo {author} {\bibfnamefont {A.~M.}\ \bibnamefont {Saxe}}, \bibinfo
  {author} {\bibfnamefont {F.}~\bibnamefont {Krzakala}},\ and\ \bibinfo
  {author} {\bibfnamefont {L.}~\bibnamefont {Zdeborov{\'{a}}}},\ }\bibfield
  {title} {\bibinfo {title} {Dynamics of stochastic gradient descent for
  two-layer neural networks in the teacher{\textendash}student setup},\ }\href
  {https://doi.org/10.1088/1742-5468/abc61e} {\bibfield  {journal} {\bibinfo
  {journal} {Journal of Statistical Mechanics: Theory and Experiment}\ }\textbf
  {\bibinfo {volume} {2020}},\ \bibinfo {pages} {124010} (\bibinfo {year}
  {2020})}\BibitemShut {NoStop}%
\bibitem [{\citenamefont {Alberici}\ \emph {et~al.}(2020)\citenamefont
  {Alberici}, \citenamefont {Barra}, \citenamefont {Contucci},\ and\
  \citenamefont {Mingione}}]{Alberici2020}%
  \BibitemOpen
  \bibfield  {author} {\bibinfo {author} {\bibfnamefont {D.}~\bibnamefont
  {Alberici}}, \bibinfo {author} {\bibfnamefont {A.}~\bibnamefont {Barra}},
  \bibinfo {author} {\bibfnamefont {P.}~\bibnamefont {Contucci}},\ and\
  \bibinfo {author} {\bibfnamefont {E.}~\bibnamefont {Mingione}},\ }\bibfield
  {title} {\bibinfo {title} {Annealing and replica-symmetry in deep boltzmann
  machines},\ }\href {https://doi.org/10.1007/s10955-020-02495-2} {\bibfield
  {journal} {\bibinfo  {journal} {Journal of Statistical Physics}\ }\textbf
  {\bibinfo {volume} {180}},\ \bibinfo {pages} {665} (\bibinfo {year}
  {2020})}\BibitemShut {NoStop}%
\bibitem [{\citenamefont {Agliari}\ \emph {et~al.}(2018)\citenamefont
  {Agliari}, \citenamefont {Migliozzi},\ and\ \citenamefont
  {Tantari}}]{Agliari2018}%
  \BibitemOpen
  \bibfield  {author} {\bibinfo {author} {\bibfnamefont {E.}~\bibnamefont
  {Agliari}}, \bibinfo {author} {\bibfnamefont {D.}~\bibnamefont {Migliozzi}},\
  and\ \bibinfo {author} {\bibfnamefont {D.}~\bibnamefont {Tantari}},\
  }\bibfield  {title} {\bibinfo {title} {Non-convex multi-species hopfield
  models},\ }\href@noop {} {\bibfield  {journal} {\bibinfo  {journal} {J. Stat.
  Phys.}\ }\textbf {\bibinfo {volume} {172}} (\bibinfo {year}
  {2018})}\BibitemShut {NoStop}%
\bibitem [{\citenamefont {Barra}\ \emph {et~al.}(2017)\citenamefont {Barra},
  \citenamefont {Genovese}, \citenamefont {Sollich},\ and\ \citenamefont
  {Tantari}}]{Barra2017}%
  \BibitemOpen
  \bibfield  {author} {\bibinfo {author} {\bibfnamefont {A.}~\bibnamefont
  {Barra}}, \bibinfo {author} {\bibfnamefont {G.}~\bibnamefont {Genovese}},
  \bibinfo {author} {\bibfnamefont {P.}~\bibnamefont {Sollich}},\ and\ \bibinfo
  {author} {\bibfnamefont {D.}~\bibnamefont {Tantari}},\ }\bibfield  {title}
  {\bibinfo {title} {Phase transitions in restricted boltzmann machines with
  generic priors},\ }\href@noop {} {\bibfield  {journal} {\bibinfo  {journal}
  {Phys. Rev. E}\ }\textbf {\bibinfo {volume} {96}} (\bibinfo {year}
  {2017})}\BibitemShut {NoStop}%
\bibitem [{\citenamefont {Barra}\ \emph {et~al.}(2018)\citenamefont {Barra},
  \citenamefont {Genovese}, \citenamefont {Sollich},\ and\ \citenamefont
  {Tantari}}]{Barra2018}%
  \BibitemOpen
  \bibfield  {author} {\bibinfo {author} {\bibfnamefont {A.}~\bibnamefont
  {Barra}}, \bibinfo {author} {\bibfnamefont {G.}~\bibnamefont {Genovese}},
  \bibinfo {author} {\bibfnamefont {P.}~\bibnamefont {Sollich}},\ and\ \bibinfo
  {author} {\bibfnamefont {D.}~\bibnamefont {Tantari}},\ }\bibfield  {title}
  {\bibinfo {title} {Phase diagram of restricted boltzmann machines and
  generalized hopfield networks with arbitrary priors},\ }\href@noop {}
  {\bibfield  {journal} {\bibinfo  {journal} {Phys. Rev. E}\ }\textbf {\bibinfo
  {volume} {97}} (\bibinfo {year} {2018})}\BibitemShut {NoStop}%
\bibitem [{\citenamefont {M{\'e}zard}(2017)}]{Mezard2017}%
  \BibitemOpen
  \bibfield  {author} {\bibinfo {author} {\bibfnamefont {M.}~\bibnamefont
  {M{\'e}zard}},\ }\bibfield  {title} {\bibinfo {title} {Mean-field
  message-passing equations in the hopfield model and its generalizations},\
  }\href@noop {} {\bibfield  {journal} {\bibinfo  {journal} {Phys. Rev. E}\
  }\textbf {\bibinfo {volume} {95}} (\bibinfo {year} {2017})}\BibitemShut
  {NoStop}%
\bibitem [{\citenamefont {Barra}\ \emph {et~al.}(2012)\citenamefont {Barra},
  \citenamefont {Bernacchia}, \citenamefont {Santucci},\ and\ \citenamefont
  {Contucci}}]{Barra2012}%
  \BibitemOpen
  \bibfield  {author} {\bibinfo {author} {\bibfnamefont {A.}~\bibnamefont
  {Barra}}, \bibinfo {author} {\bibfnamefont {A.}~\bibnamefont {Bernacchia}},
  \bibinfo {author} {\bibfnamefont {E.}~\bibnamefont {Santucci}},\ and\
  \bibinfo {author} {\bibfnamefont {P.}~\bibnamefont {Contucci}},\ }\bibfield
  {title} {\bibinfo {title} {On the equivalence of hopfield networks and
  boltzmann machines},\ }\href@noop {} {\bibfield  {journal} {\bibinfo
  {journal} {Neural Netw.}\ }\textbf {\bibinfo {volume} {34}} (\bibinfo {year}
  {2012})}\BibitemShut {NoStop}%
\bibitem [{\citenamefont {M{\'e}zard}\ \emph {et~al.}(1987)\citenamefont
  {M{\'e}zard}, \citenamefont {Parisi},\ and\ \citenamefont
  {Virasoro}}]{Mezard1987}%
  \BibitemOpen
  \bibfield  {author} {\bibinfo {author} {\bibfnamefont {M.}~\bibnamefont
  {M{\'e}zard}}, \bibinfo {author} {\bibfnamefont {G.}~\bibnamefont {Parisi}},\
  and\ \bibinfo {author} {\bibfnamefont {M.~A.}\ \bibnamefont {Virasoro}},\
  }\href@noop {} {\emph {\bibinfo {title} {Spin Glass Theory and Beyond: An
  Introduction to the Replica Method and Its Applications}}}\ (\bibinfo
  {publisher} {World Scientific},\ \bibinfo {address} {Singapore},\ \bibinfo
  {year} {1987})\BibitemShut {NoStop}%
\bibitem [{\citenamefont {Halverson}\ \emph {et~al.}(2020)\citenamefont
  {Halverson}, \citenamefont {Maiti},\ and\ \citenamefont
  {Stoner}}]{Halverson:2020trp}%
  \BibitemOpen
  \bibfield  {author} {\bibinfo {author} {\bibfnamefont {J.}~\bibnamefont
  {Halverson}}, \bibinfo {author} {\bibfnamefont {A.}~\bibnamefont {Maiti}},\
  and\ \bibinfo {author} {\bibfnamefont {K.}~\bibnamefont {Stoner}},\
  }\bibfield  {title} {\bibinfo {title} {{Neural Networks and Quantum Field
  Theory}},\ }\href@noop {} {\  (\bibinfo {year} {2020})},\ \Eprint
  {https://arxiv.org/abs/2008.08601} {arXiv:2008.08601 [cs.LG]} \BibitemShut
  {NoStop}%
\bibitem [{\citenamefont {Lee}\ \emph {et~al.}(2018)\citenamefont {Lee},
  \citenamefont {Bahri}, \citenamefont {Novak}, \citenamefont {Schoenholz},
  \citenamefont {Pennington},\ and\ \citenamefont {Sohl-dickstein}}]{46760}%
  \BibitemOpen
  \bibfield  {author} {\bibinfo {author} {\bibfnamefont {J.}~\bibnamefont
  {Lee}}, \bibinfo {author} {\bibfnamefont {Y.}~\bibnamefont {Bahri}}, \bibinfo
  {author} {\bibfnamefont {R.}~\bibnamefont {Novak}}, \bibinfo {author}
  {\bibfnamefont {S.}~\bibnamefont {Schoenholz}}, \bibinfo {author}
  {\bibfnamefont {J.}~\bibnamefont {Pennington}},\ and\ \bibinfo {author}
  {\bibfnamefont {J.}~\bibnamefont {Sohl-dickstein}},\ }\bibfield  {title}
  {\bibinfo {title} {Deep neural networks as gaussian processes}\ }(\bibinfo
  {address} {6th International Conference on Learning Representations,
  Vancouver, BC, Canada},\ \bibinfo {year} {2018})\BibitemShut {NoStop}%
\bibitem [{\citenamefont {de~G.~Matthews}\ \emph {et~al.}(2018)\citenamefont
  {de~G.~Matthews}, \citenamefont {Hron}, \citenamefont {Rowland},
  \citenamefont {Turner},\ and\ \citenamefont {Ghahramani}}]{g.2018gaussian}%
  \BibitemOpen
  \bibfield  {author} {\bibinfo {author} {\bibfnamefont {A.~G.}\ \bibnamefont
  {de~G.~Matthews}}, \bibinfo {author} {\bibfnamefont {J.}~\bibnamefont
  {Hron}}, \bibinfo {author} {\bibfnamefont {M.}~\bibnamefont {Rowland}},
  \bibinfo {author} {\bibfnamefont {R.~E.}\ \bibnamefont {Turner}},\ and\
  \bibinfo {author} {\bibfnamefont {Z.}~\bibnamefont {Ghahramani}},\ }\bibfield
   {title} {\bibinfo {title} {Gaussian process behaviour in wide deep neural
  networks},\ }in\ \href {https://openreview.net/forum?id=H1-nGgWC-} {\emph
  {\bibinfo {booktitle} {International Conference on Learning
  Representations}}}\ (\bibinfo {year} {2018})\BibitemShut {NoStop}%
\bibitem [{\citenamefont {Novak}\ \emph {et~al.}(2019)\citenamefont {Novak},
  \citenamefont {Xiao}, \citenamefont {Bahri}, \citenamefont {Lee},
  \citenamefont {Yang}, \citenamefont {Abolafia}, \citenamefont {Pennington},\
  and\ \citenamefont {Sohl-dickstein}}]{novak2019bayesian}%
  \BibitemOpen
  \bibfield  {author} {\bibinfo {author} {\bibfnamefont {R.}~\bibnamefont
  {Novak}}, \bibinfo {author} {\bibfnamefont {L.}~\bibnamefont {Xiao}},
  \bibinfo {author} {\bibfnamefont {Y.}~\bibnamefont {Bahri}}, \bibinfo
  {author} {\bibfnamefont {J.}~\bibnamefont {Lee}}, \bibinfo {author}
  {\bibfnamefont {G.}~\bibnamefont {Yang}}, \bibinfo {author} {\bibfnamefont
  {D.~A.}\ \bibnamefont {Abolafia}}, \bibinfo {author} {\bibfnamefont
  {J.}~\bibnamefont {Pennington}},\ and\ \bibinfo {author} {\bibfnamefont
  {J.}~\bibnamefont {Sohl-dickstein}},\ }\bibfield  {title} {\bibinfo {title}
  {Bayesian deep convolutional networks with many channels are gaussian
  processes},\ }in\ \href {https://openreview.net/forum?id=B1g30j0qF7} {\emph
  {\bibinfo {booktitle} {International Conference on Learning
  Representations}}}\ (\bibinfo {year} {2019})\BibitemShut {NoStop}%
\bibitem [{\citenamefont {Garriga-Alonso}\ \emph {et~al.}(2019)\citenamefont
  {Garriga-Alonso}, \citenamefont {Rasmussen},\ and\ \citenamefont
  {Aitchison}}]{garriga-alonso2018deep}%
  \BibitemOpen
  \bibfield  {author} {\bibinfo {author} {\bibfnamefont {A.}~\bibnamefont
  {Garriga-Alonso}}, \bibinfo {author} {\bibfnamefont {C.~E.}\ \bibnamefont
  {Rasmussen}},\ and\ \bibinfo {author} {\bibfnamefont {L.}~\bibnamefont
  {Aitchison}},\ }\bibfield  {title} {\bibinfo {title} {Deep convolutional
  networks as shallow gaussian processes},\ }in\ \href
  {https://openreview.net/forum?id=Bklfsi0cKm} {\emph {\bibinfo {booktitle}
  {International Conference on Learning Representations}}}\ (\bibinfo {year}
  {2019})\BibitemShut {NoStop}%
\bibitem [{\citenamefont {Koller}\ and\ \citenamefont
  {Friedman}(2009)}]{Koller}%
  \BibitemOpen
  \bibfield  {author} {\bibinfo {author} {\bibfnamefont {D.}~\bibnamefont
  {Koller}}\ and\ \bibinfo {author} {\bibfnamefont {N.}~\bibnamefont
  {Friedman}},\ }\href@noop {} {\emph {\bibinfo {title} {Probabilistic
  Graphical Models: Principles and Techniques}}}\ (\bibinfo  {publisher} {The
  MIT Press},\ \bibinfo {year} {2009})\BibitemShut {NoStop}%
\bibitem [{\citenamefont {Wilson}(1974)}]{PhysRevD.10.2445}%
  \BibitemOpen
  \bibfield  {author} {\bibinfo {author} {\bibfnamefont {K.~G.}\ \bibnamefont
  {Wilson}},\ }\bibfield  {title} {\bibinfo {title} {Confinement of quarks},\
  }\href {https://doi.org/10.1103/PhysRevD.10.2445} {\bibfield  {journal}
  {\bibinfo  {journal} {Phys. Rev. D}\ }\textbf {\bibinfo {volume} {10}},\
  \bibinfo {pages} {2445} (\bibinfo {year} {1974})}\BibitemShut {NoStop}%
\bibitem [{\citenamefont {Preston}(1974)}]{Preston}%
  \BibitemOpen
  \bibfield  {author} {\bibinfo {author} {\bibfnamefont {C.~J.}\ \bibnamefont
  {Preston}},\ }\href {https://doi.org/10.1017/CBO9780511897122} {\emph
  {\bibinfo {title} {Gibbs States on Countable Sets}}},\ Cambridge Tracts in
  Mathematics\ (\bibinfo  {publisher} {Cambridge University Press},\ \bibinfo
  {year} {1974})\BibitemShut {NoStop}%
\bibitem [{\citenamefont {Milchev}\ \emph {et~al.}(1986)\citenamefont
  {Milchev}, \citenamefont {Heermann},\ and\ \citenamefont
  {Binder}}]{Milchev1986}%
  \BibitemOpen
  \bibfield  {author} {\bibinfo {author} {\bibfnamefont {A.}~\bibnamefont
  {Milchev}}, \bibinfo {author} {\bibfnamefont {D.~W.}\ \bibnamefont
  {Heermann}},\ and\ \bibinfo {author} {\bibfnamefont {K.}~\bibnamefont
  {Binder}},\ }\bibfield  {title} {\bibinfo {title} {Finite-size scaling
  analysis of the $\phi$4 field theory on the square lattice},\ }\href
  {https://doi.org/10.1007/BF01011906} {\bibfield  {journal} {\bibinfo
  {journal} {Journal of Statistical Physics}\ }\textbf {\bibinfo {volume}
  {44}},\ \bibinfo {pages} {749} (\bibinfo {year} {1986})}\BibitemShut
  {NoStop}%
\bibitem [{\citenamefont {Bishop}(2006)}]{10.5555/1162264}%
  \BibitemOpen
  \bibfield  {author} {\bibinfo {author} {\bibfnamefont {C.~M.}\ \bibnamefont
  {Bishop}},\ }\href@noop {} {\emph {\bibinfo {title} {Pattern Recognition and
  Machine Learning (Information Science and Statistics)}}}\ (\bibinfo
  {publisher} {Springer-Verlag},\ \bibinfo {address} {Berlin, Heidelberg},\
  \bibinfo {year} {2006})\BibitemShut {NoStop}%
\bibitem [{\citenamefont {Ferrenberg}\ and\ \citenamefont
  {Swendsen}(1988)}]{PhysRevLett.61.2635}%
  \BibitemOpen
  \bibfield  {author} {\bibinfo {author} {\bibfnamefont {A.~M.}\ \bibnamefont
  {Ferrenberg}}\ and\ \bibinfo {author} {\bibfnamefont {R.~H.}\ \bibnamefont
  {Swendsen}},\ }\bibfield  {title} {\bibinfo {title} {New monte carlo
  technique for studying phase transitions},\ }\href
  {https://doi.org/10.1103/PhysRevLett.61.2635} {\bibfield  {journal} {\bibinfo
   {journal} {Phys. Rev. Lett.}\ }\textbf {\bibinfo {volume} {61}},\ \bibinfo
  {pages} {2635} (\bibinfo {year} {1988})}\BibitemShut {NoStop}%
\bibitem [{\citenamefont {Blake}\ \emph {et~al.}(2011)\citenamefont {Blake},
  \citenamefont {Kohli},\ and\ \citenamefont {Rother}}]{10.5555/2024611}%
  \BibitemOpen
  \bibfield  {author} {\bibinfo {author} {\bibfnamefont {A.}~\bibnamefont
  {Blake}}, \bibinfo {author} {\bibfnamefont {P.}~\bibnamefont {Kohli}},\ and\
  \bibinfo {author} {\bibfnamefont {C.}~\bibnamefont {Rother}},\ }\href@noop {}
  {\emph {\bibinfo {title} {Markov Random Fields for Vision and Image
  Processing}}}\ (\bibinfo  {publisher} {The MIT Press},\ \bibinfo {year}
  {2011})\BibitemShut {NoStop}%
\bibitem [{\citenamefont {Krizhevsky}(2009)}]{Krizhevsky2009LearningML}%
  \BibitemOpen
  \bibfield  {author} {\bibinfo {author} {\bibfnamefont {A.}~\bibnamefont
  {Krizhevsky}},\ }\bibfield  {title} {\bibinfo {title} {Learning multiple
  layers of features from tiny images}\ }(\bibinfo {year} {2009})\BibitemShut
  {NoStop}%
\bibitem [{\citenamefont {Smolensky}(1986)}]{10.5555/104279.104290}%
  \BibitemOpen
  \bibfield  {author} {\bibinfo {author} {\bibfnamefont {P.}~\bibnamefont
  {Smolensky}},\ }\bibinfo {title} {Information processing in dynamical
  systems: Foundations of harmony theory},\ in\ \href@noop {} {\emph {\bibinfo
  {booktitle} {Parallel Distributed Processing: Explorations in the
  Microstructure of Cognition, Vol. 1: Foundations}}}\ (\bibinfo  {publisher}
  {MIT Press},\ \bibinfo {address} {Cambridge, MA, USA},\ \bibinfo {year}
  {1986})\ p.\ \bibinfo {pages} {194–281}\BibitemShut {NoStop}%
\bibitem [{\citenamefont {Ackley}\ \emph {et~al.}(1985)\citenamefont {Ackley},
  \citenamefont {Hinton},\ and\ \citenamefont {Sejnowski}}]{ACKLEY1985147}%
  \BibitemOpen
  \bibfield  {author} {\bibinfo {author} {\bibfnamefont {D.~H.}\ \bibnamefont
  {Ackley}}, \bibinfo {author} {\bibfnamefont {G.~E.}\ \bibnamefont {Hinton}},\
  and\ \bibinfo {author} {\bibfnamefont {T.~J.}\ \bibnamefont {Sejnowski}},\
  }\bibfield  {title} {\bibinfo {title} {A learning algorithm for boltzmann
  machines},\ }\href
  {https://doi.org/https://doi.org/10.1016/S0364-0213(85)80012-4} {\bibfield
  {journal} {\bibinfo  {journal} {Cognitive Science}\ }\textbf {\bibinfo
  {volume} {9}},\ \bibinfo {pages} {147 } (\bibinfo {year} {1985})}\BibitemShut
  {NoStop}%
\bibitem [{\citenamefont {Fischer}\ and\ \citenamefont
  {Igel}(2014)}]{FISCHER201425}%
  \BibitemOpen
  \bibfield  {author} {\bibinfo {author} {\bibfnamefont {A.}~\bibnamefont
  {Fischer}}\ and\ \bibinfo {author} {\bibfnamefont {C.}~\bibnamefont {Igel}},\
  }\bibfield  {title} {\bibinfo {title} {Training restricted boltzmann
  machines: An introduction},\ }\href
  {https://doi.org/https://doi.org/10.1016/j.patcog.2013.05.025} {\bibfield
  {journal} {\bibinfo  {journal} {Pattern Recognition}\ }\textbf {\bibinfo
  {volume} {47}},\ \bibinfo {pages} {25 } (\bibinfo {year} {2014})}\BibitemShut
  {NoStop}%
\bibitem [{\citenamefont {Hinton}(2012)}]{Hinton2012}%
  \BibitemOpen
  \bibfield  {author} {\bibinfo {author} {\bibfnamefont {G.~E.}\ \bibnamefont
  {Hinton}},\ }\bibinfo {title} {A practical guide to training restricted
  boltzmann machines},\ in\ \href
  {https://doi.org/10.1007/978-3-642-35289-8_32} {\emph {\bibinfo {booktitle}
  {Neural Networks: Tricks of the Trade: Second Edition}}},\ \bibinfo {editor}
  {edited by\ \bibinfo {editor} {\bibfnamefont {G.}~\bibnamefont {Montavon}},
  \bibinfo {editor} {\bibfnamefont {G.~B.}\ \bibnamefont {Orr}},\ and\ \bibinfo
  {editor} {\bibfnamefont {K.-R.}\ \bibnamefont {M{\"u}ller}}}\ (\bibinfo
  {publisher} {Springer Berlin Heidelberg},\ \bibinfo {address} {Berlin,
  Heidelberg},\ \bibinfo {year} {2012})\ pp.\ \bibinfo {pages}
  {599--619}\BibitemShut {NoStop}%
\bibitem [{\citenamefont {Bengio}\ \emph {et~al.}(2006)\citenamefont {Bengio},
  \citenamefont {Lamblin}, \citenamefont {Popovici},\ and\ \citenamefont
  {Larochelle}}]{10.5555/2976456.2976476}%
  \BibitemOpen
  \bibfield  {author} {\bibinfo {author} {\bibfnamefont {Y.}~\bibnamefont
  {Bengio}}, \bibinfo {author} {\bibfnamefont {P.}~\bibnamefont {Lamblin}},
  \bibinfo {author} {\bibfnamefont {D.}~\bibnamefont {Popovici}},\ and\
  \bibinfo {author} {\bibfnamefont {H.}~\bibnamefont {Larochelle}},\ }\bibfield
   {title} {\bibinfo {title} {Greedy layer-wise training of deep networks},\
  }in\ \href@noop {} {\emph {\bibinfo {booktitle} {Proceedings of the 19th
  International Conference on Neural Information Processing Systems}}},\
  \bibinfo {series and number} {NIPS'06}\ (\bibinfo  {publisher} {MIT Press},\
  \bibinfo {address} {Cambridge, MA, USA},\ \bibinfo {year} {2006})\ p.\
  \bibinfo {pages} {153–160}\BibitemShut {NoStop}%
\bibitem [{\citenamefont {Krause}\ \emph {et~al.}(2013)\citenamefont {Krause},
  \citenamefont {Fischer}, \citenamefont {Glasmachers},\ and\ \citenamefont
  {Igel}}]{pmlr-v28-krause13}%
  \BibitemOpen
  \bibfield  {author} {\bibinfo {author} {\bibfnamefont {O.}~\bibnamefont
  {Krause}}, \bibinfo {author} {\bibfnamefont {A.}~\bibnamefont {Fischer}},
  \bibinfo {author} {\bibfnamefont {T.}~\bibnamefont {Glasmachers}},\ and\
  \bibinfo {author} {\bibfnamefont {C.}~\bibnamefont {Igel}},\ }\bibfield
  {title} {\bibinfo {title} {Approximation properties of {DBNs} with binary
  hidden units and real-valued visible units},\ }in\ \href
  {http://proceedings.mlr.press/v28/krause13.html} {\emph {\bibinfo {booktitle}
  {Proceedings of the 30th International Conference on Machine Learning}}},\
  \bibinfo {series} {Proceedings of Machine Learning Research}, Vol.~\bibinfo
  {volume} {28},\ \bibinfo {editor} {edited by\ \bibinfo {editor}
  {\bibfnamefont {S.}~\bibnamefont {Dasgupta}}\ and\ \bibinfo {editor}
  {\bibfnamefont {D.}~\bibnamefont {McAllester}}}\ (\bibinfo  {publisher}
  {PMLR},\ \bibinfo {address} {Atlanta, Georgia, USA},\ \bibinfo {year}
  {2013})\ pp.\ \bibinfo {pages} {419--426}\BibitemShut {NoStop}%
\bibitem [{Note1()}]{Note1}%
  \BibitemOpen
  \bibinfo {note} {This data set contains a set of face images taken between
  April 1992 and April 1994 at AT\&T Laboratories Cambridge}\BibitemShut
  {NoStop}%
\bibitem [{\citenamefont {Hinton}\ and\ \citenamefont
  {Salakhutdinov}(2006)}]{Hinton504}%
  \BibitemOpen
  \bibfield  {author} {\bibinfo {author} {\bibfnamefont {G.~E.}\ \bibnamefont
  {Hinton}}\ and\ \bibinfo {author} {\bibfnamefont {R.~R.}\ \bibnamefont
  {Salakhutdinov}},\ }\bibfield  {title} {\bibinfo {title} {Reducing the
  dimensionality of data with neural networks},\ }\href
  {https://doi.org/10.1126/science.1127647} {\bibfield  {journal} {\bibinfo
  {journal} {Science}\ }\textbf {\bibinfo {volume} {313}},\ \bibinfo {pages}
  {504} (\bibinfo {year} {2006})}\BibitemShut {NoStop}%
\bibitem [{\citenamefont {Hands}(1988)}]{HANDS1988597}%
  \BibitemOpen
  \bibfield  {author} {\bibinfo {author} {\bibfnamefont {S.}~\bibnamefont
  {Hands}},\ }\bibfield  {title} {\bibinfo {title} {Abelian gauge glasses},\
  }\href {https://doi.org/https://doi.org/10.1016/0550-3213(88)90118-6}
  {\bibfield  {journal} {\bibinfo  {journal} {Nuclear Physics B}\ }\textbf
  {\bibinfo {volume} {305}},\ \bibinfo {pages} {597} (\bibinfo {year}
  {1988})}\BibitemShut {NoStop}%
\bibitem [{\citenamefont {Narovlansky}\ and\ \citenamefont
  {Aharony}(2018)}]{PhysRevLett.121.071601}%
  \BibitemOpen
  \bibfield  {author} {\bibinfo {author} {\bibfnamefont {V.}~\bibnamefont
  {Narovlansky}}\ and\ \bibinfo {author} {\bibfnamefont {O.}~\bibnamefont
  {Aharony}},\ }\bibfield  {title} {\bibinfo {title} {Renormalization group in
  field theories with quantum quenched disorder},\ }\href
  {https://doi.org/10.1103/PhysRevLett.121.071601} {\bibfield  {journal}
  {\bibinfo  {journal} {Phys. Rev. Lett.}\ }\textbf {\bibinfo {volume} {121}},\
  \bibinfo {pages} {071601} (\bibinfo {year} {2018})}\BibitemShut {NoStop}%
\bibitem [{\citenamefont {Aharony}\ and\ \citenamefont
  {Narovlansky}(2018)}]{PhysRevD.98.045012}%
  \BibitemOpen
  \bibfield  {author} {\bibinfo {author} {\bibfnamefont {O.}~\bibnamefont
  {Aharony}}\ and\ \bibinfo {author} {\bibfnamefont {V.}~\bibnamefont
  {Narovlansky}},\ }\bibfield  {title} {\bibinfo {title} {Renormalization group
  flow in field theories with quenched disorder},\ }\href
  {https://doi.org/10.1103/PhysRevD.98.045012} {\bibfield  {journal} {\bibinfo
  {journal} {Phys. Rev. D}\ }\textbf {\bibinfo {volume} {98}},\ \bibinfo
  {pages} {045012} (\bibinfo {year} {2018})}\BibitemShut {NoStop}%
\bibitem [{\citenamefont {Marinari}\ and\ \citenamefont
  {Parisi}(1992)}]{Marinari_1992}%
  \BibitemOpen
  \bibfield  {author} {\bibinfo {author} {\bibfnamefont {E.}~\bibnamefont
  {Marinari}}\ and\ \bibinfo {author} {\bibfnamefont {G.}~\bibnamefont
  {Parisi}},\ }\bibfield  {title} {\bibinfo {title} {Simulated tempering: A new
  monte carlo scheme},\ }\href {https://doi.org/10.1209/0295-5075/19/6/002}
  {\bibfield  {journal} {\bibinfo  {journal} {Europhysics Letters ({EPL})}\
  }\textbf {\bibinfo {volume} {19}},\ \bibinfo {pages} {451} (\bibinfo {year}
  {1992})}\BibitemShut {NoStop}%
\bibitem [{\citenamefont {Lee}(1993)}]{PhysRevLett.71.211}%
  \BibitemOpen
  \bibfield  {author} {\bibinfo {author} {\bibfnamefont {J.}~\bibnamefont
  {Lee}},\ }\bibfield  {title} {\bibinfo {title} {New monte carlo algorithm:
  Entropic sampling},\ }\href {https://doi.org/10.1103/PhysRevLett.71.211}
  {\bibfield  {journal} {\bibinfo  {journal} {Phys. Rev. Lett.}\ }\textbf
  {\bibinfo {volume} {71}},\ \bibinfo {pages} {211} (\bibinfo {year}
  {1993})}\BibitemShut {NoStop}%
\bibitem [{\citenamefont {Brower}\ and\ \citenamefont
  {Tamayo}(1989)}]{PhysRevLett.62.1087}%
  \BibitemOpen
  \bibfield  {author} {\bibinfo {author} {\bibfnamefont {R.~C.}\ \bibnamefont
  {Brower}}\ and\ \bibinfo {author} {\bibfnamefont {P.}~\bibnamefont
  {Tamayo}},\ }\bibfield  {title} {\bibinfo {title} {Embedded dynamics for
  ${\mathrm{\ensuremath{\varphi}}}^{4}$ theory},\ }\href
  {https://doi.org/10.1103/PhysRevLett.62.1087} {\bibfield  {journal} {\bibinfo
   {journal} {Phys. Rev. Lett.}\ }\textbf {\bibinfo {volume} {62}},\ \bibinfo
  {pages} {1087} (\bibinfo {year} {1989})}\BibitemShut {NoStop}%
\bibitem [{\citenamefont {Loinaz}\ and\ \citenamefont
  {Willey}(1998)}]{PhysRevD.58.076003}%
  \BibitemOpen
  \bibfield  {author} {\bibinfo {author} {\bibfnamefont {W.}~\bibnamefont
  {Loinaz}}\ and\ \bibinfo {author} {\bibfnamefont {R.~S.}\ \bibnamefont
  {Willey}},\ }\bibfield  {title} {\bibinfo {title} {Monte carlo simulation
  calculation of the critical coupling constant for two-dimensional continuum
  ${\ensuremath{\varphi}}^{4}$ theory},\ }\href
  {https://doi.org/10.1103/PhysRevD.58.076003} {\bibfield  {journal} {\bibinfo
  {journal} {Phys. Rev. D}\ }\textbf {\bibinfo {volume} {58}},\ \bibinfo
  {pages} {076003} (\bibinfo {year} {1998})}\BibitemShut {NoStop}%
\bibitem [{\citenamefont {Bietenholz}\ \emph {et~al.}(1997)\citenamefont
  {Bietenholz}, \citenamefont {Brower}, \citenamefont {Chandrasekharan},\ and\
  \citenamefont {Wiese}}]{BIETENHOLZ1997921}%
  \BibitemOpen
  \bibfield  {author} {\bibinfo {author} {\bibfnamefont {W.}~\bibnamefont
  {Bietenholz}}, \bibinfo {author} {\bibfnamefont {R.}~\bibnamefont {Brower}},
  \bibinfo {author} {\bibfnamefont {S.}~\bibnamefont {Chandrasekharan}},\ and\
  \bibinfo {author} {\bibfnamefont {U.-J.}\ \bibnamefont {Wiese}},\ }\bibfield
  {title} {\bibinfo {title} {Progress on perfect lattice actions for qcd},\
  }\href {https://doi.org/https://doi.org/10.1016/S0920-5632(96)00818-3}
  {\bibfield  {journal} {\bibinfo  {journal} {Nuclear Physics B - Proceedings
  Supplements}\ }\textbf {\bibinfo {volume} {53}},\ \bibinfo {pages} {921}
  (\bibinfo {year} {1997})},\ \bibinfo {note} {lattice 96}\BibitemShut
  {NoStop}%
\bibitem [{\citenamefont {Bietenholz}\ and\ \citenamefont
  {Wiese}(1998)}]{BIETENHOLZ1998114}%
  \BibitemOpen
  \bibfield  {author} {\bibinfo {author} {\bibfnamefont {W.}~\bibnamefont
  {Bietenholz}}\ and\ \bibinfo {author} {\bibfnamefont {U.-J.}\ \bibnamefont
  {Wiese}},\ }\bibfield  {title} {\bibinfo {title} {Perfect actions with
  chemical potential},\ }\href
  {https://doi.org/https://doi.org/10.1016/S0370-2693(98)00269-X} {\bibfield
  {journal} {\bibinfo  {journal} {Physics Letters B}\ }\textbf {\bibinfo
  {volume} {426}},\ \bibinfo {pages} {114} (\bibinfo {year}
  {1998})}\BibitemShut {NoStop}%
\bibitem [{\citenamefont {Jain}\ and\ \citenamefont
  {Vanchurin}(2016)}]{Jain2016}%
  \BibitemOpen
  \bibfield  {author} {\bibinfo {author} {\bibfnamefont {M.}~\bibnamefont
  {Jain}}\ and\ \bibinfo {author} {\bibfnamefont {V.}~\bibnamefont
  {Vanchurin}},\ }\bibfield  {title} {\bibinfo {title} {Generating functionals
  for quantum field theories with random potentials},\ }\href
  {https://doi.org/10.1007/JHEP01(2016)107} {\bibfield  {journal} {\bibinfo
  {journal} {Journal of High Energy Physics}\ }\textbf {\bibinfo {volume}
  {2016}},\ \bibinfo {pages} {107} (\bibinfo {year} {2016})}\BibitemShut
  {NoStop}%
\end{thebibliography}%
\end{document}